\documentclass[graybox, envcountchap]{svmult}

\usepackage{mathptmx}        
\usepackage{helvet}          
\usepackage{courier}         

\usepackage{makeidx}         
\usepackage{graphicx}        
\usepackage{multicol}        
\usepackage[bottom]{footmisc}

\makeindex             

\usepackage{amsmath}
\usepackage{amssymb}
\usepackage{lscape}
\usepackage{multirow}

\def\gapp{\ifmmode\stackrel{>}{_{\sim}}\else$\stackrel{<}{_{\sim}}$\fi}
\def\gsim{\lower.5ex\hbox{\gtsima}}
\def\gtsima{$\; \buildrel > \over \sim \;$}
\def\lapp{\ifmmode\stackrel{<}{_{\sim}}\else$\stackrel{<}{_{\sim}}$\fi}
\def\lsim{\lower.5ex\hbox{\ltsima}}
\def\ltsima{$\; \buildrel < \over \sim \;$}

\newcommand\apgt{\ {\raise-.5ex\hbox{$\buildrel>\over\sim$}}\ }
\newcommand\aplt{\ {\raise-.5ex\hbox{$\buildrel<\over\sim$}}\ }
\begin{document}
\pagestyle{empty}
\frontmatter

\include{dedic}
\include{foreword}
\include{preface}

\mainmatter

\setcounter{chapter}{13}

\title{Blue Stragglers in Clusters and Integrated Spectral Properties of Stellar Populations}
\titlerunning{Blue Stragglers and ISEDs of SSPs}
\author{Yu Xin \& Licai Deng}
\institute{Yu Xin \& Licai Deng \at National Astronomical Observatories, Chinese Academy of Sciences \email{xinyu@bao.ac.cn, licai@bao.ac.cn}
}
%
%
\maketitle
\label{Chapter:Deng}

\abstract*{Blue straggler stars are the most prominent bright objects in the colour-magnitude diagram of a star cluster that challenges the theory of stellar evolution. Star clusters are the closest counterparts of the theoretical concept of simple 
stellar populations (SSPs) in the Universe. SSPs are widely used as the basic building blocks
to interpret stellar contents in galaxies. The concept of an SSP is a group of coeval stars 
which follows a given distribution in mass, and has the same chemical property and age. 
In practice, SSPs are more conveniently made by the latest stellar evolutionary models of 
single stars. In reality, however, stars can be more complicated than just single either at 
birth time or during the course of evolution in a typical environment. Observations of star 
clusters show that there are always exotic objects which do not follow the predictions of 
standard theory of stellar evolution. Blue straggler stars (BSSs), as discussed intensively in 
this book both observationally and theoretically, are very important in our context when 
considering the integrated spectral properties of a cluster, or a simple stellar population. 
In this chapter, we are going to describe how important the contribution of BSSs is
to the total light of a cluster.}

\section{Introduction}
\label{sec:1}\sectionmark{Introduction}
Evolutionary population synthesis\index{population synthesis} (EPS) has been widely used as a powerful tool to study 
the stellar contents of galaxies\index{galaxy}. In essence,
EPS compares the observed, integrated spectrum\index{integrated spectrum} of a galaxy with a combination of spectra 
of simple stellar populations\index{stellar population} (SSPs;
single-age, single-metallicity populations) of different ages\index{age} and metallicities\index{metallicity}, and decomposes
the complex stellar contents into SSPs of known ages and metallicities to infer the 
galaxy's star-formation history. Over the
past two decades, much work has been done to improve the accuracy of EPS and SSP models 
in various contexts (e.g., \cite{BA86,BC03,Leitherer+99,Thomas+03,Vazdekis99,Worthey94}). 

Unfortunately, population synthesis models still suffer from a number of limitations. 
One is our poor understanding of some advanced single-star evolutionary\index{stellar evolution} phases, 
such as of supergiants and asymptotic giant branch stars\index{asymptotic giant branch star} \cite{Yi03}, while a second is an 
absence in the models of the results of stellar interactions, such as the so-called ``stragglers'' 
formed through mass transfer\index{mass transfer} in binaries\index{binary system} or stellar collisions\index{collision}. Such stars are usually very bright 
and can strongly affect the integrated-light properties of the entire
system. The potential uncertainties inherent to EPS caused by ignoring these components could be 
much larger than those still remaining and due to the variety of input physics among different models.

In this chapter, we focus on the second limitation to the standard SSP models. 
With the updated knowledge of stars, in particular on the physical properties of binary 
and collisional interactions that eventually create the exotic blue straggler stars (BSSs)
discussed in previous chapters in this book, a more realistic prescription of stellar populations 
is now possible. The goal of this chapter is to  present a new set of SSP models which include 
contributions from BSSs.

\begin{figure}
\includegraphics[width=5.9cm]{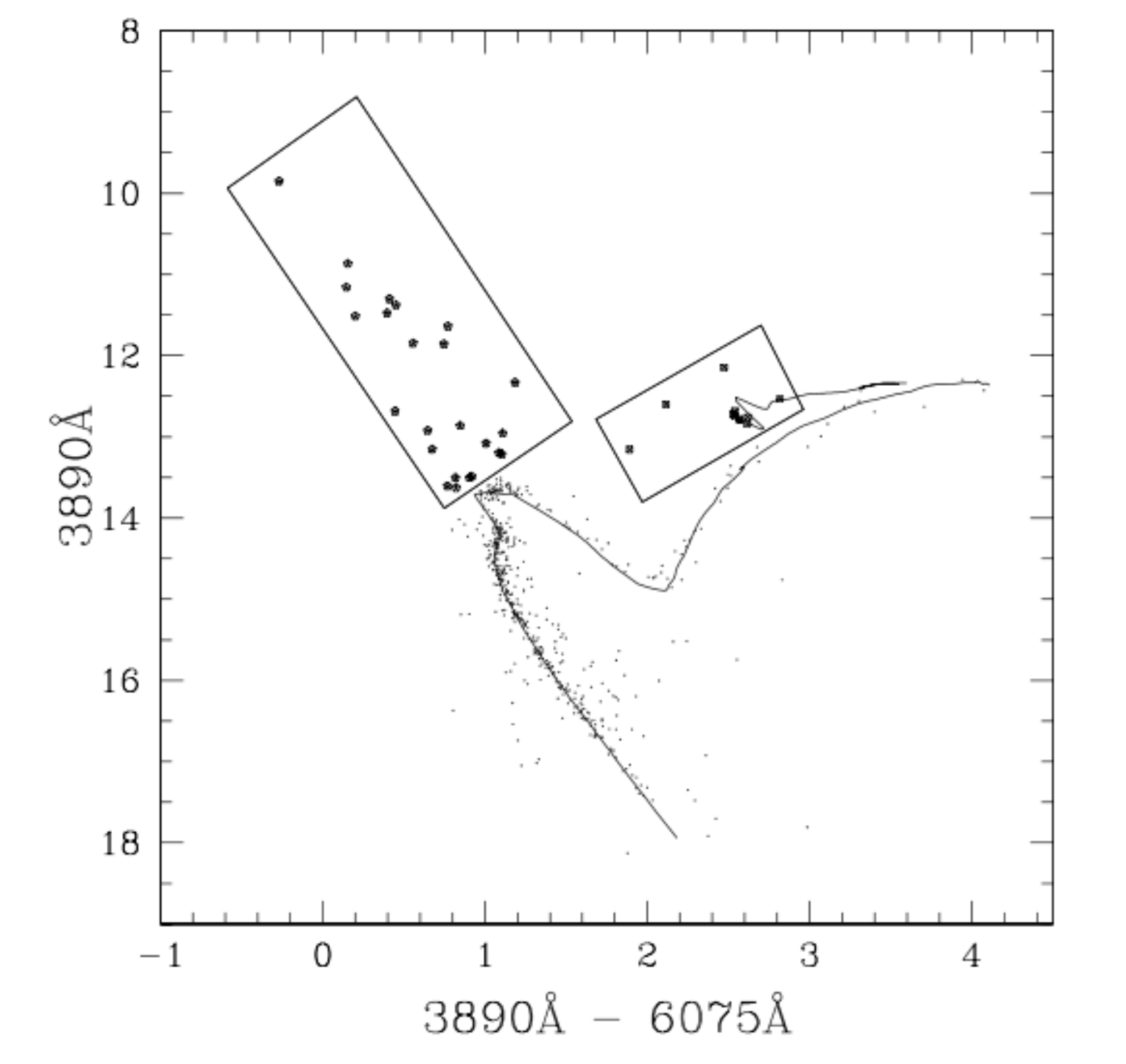}\includegraphics[width=5.9cm]{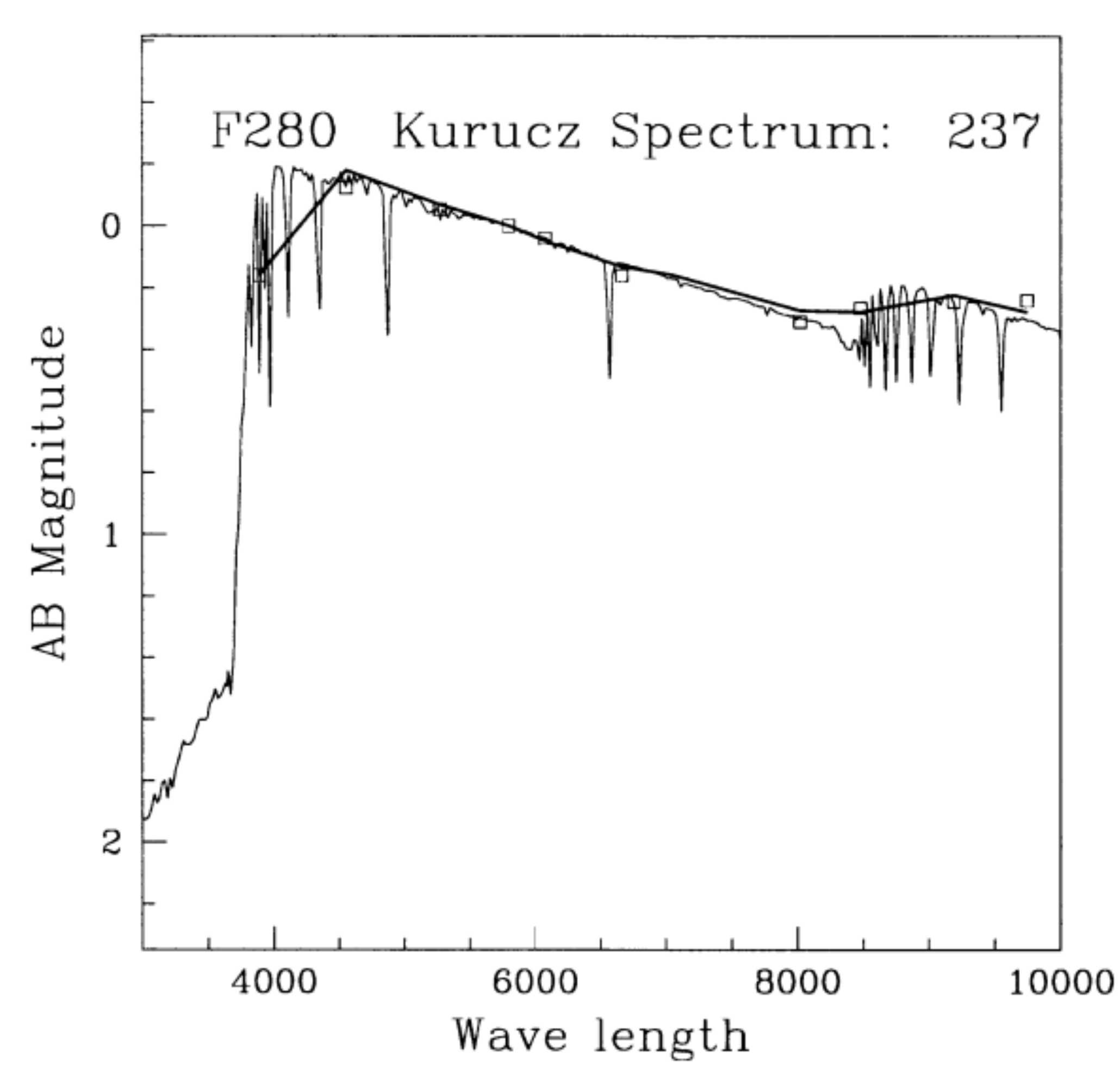}\\
\includegraphics[width=5.9cm]{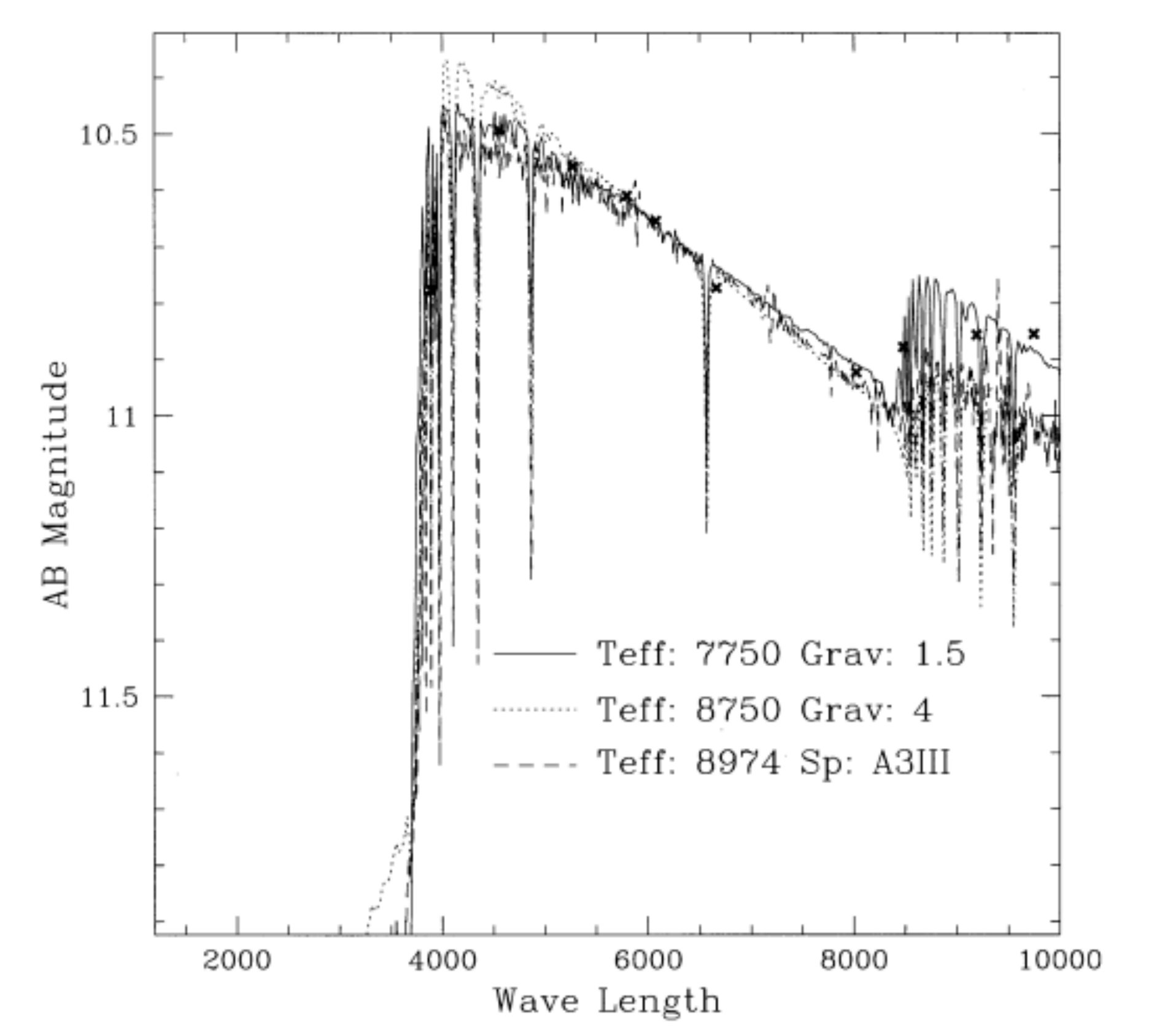}\includegraphics[width=5.9cm]{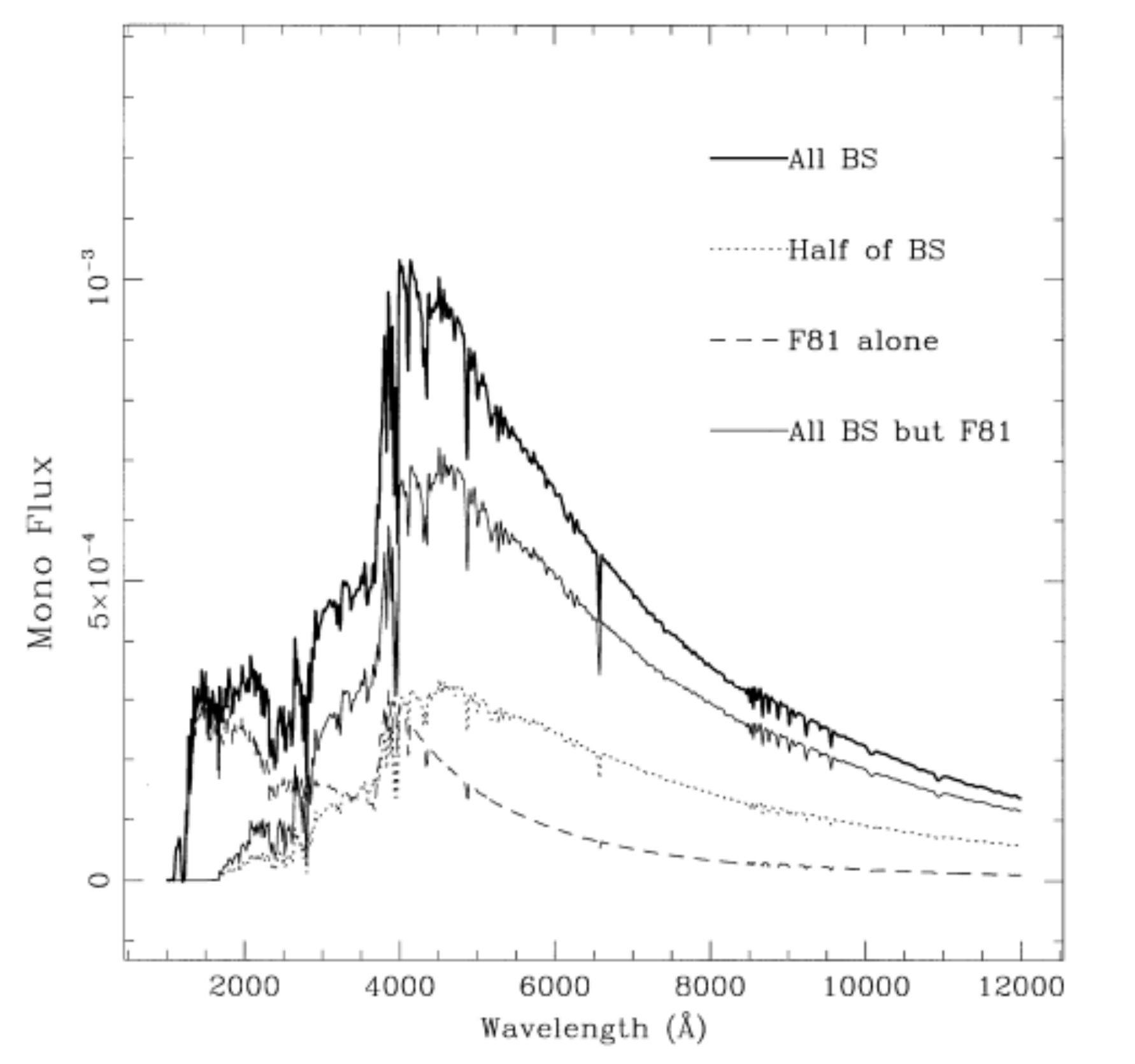}
\caption{This is an example of building up the integrated spectral energy distribution (ISED) of a star cluster based on accurate
spectrophotometry\index{spectrophotometry} from the BATC survey \cite{Fan+96}. \emph{Upper left panel}: CMD of M67\index{M67} with 
all highly probable members, the left box along the extension of the main sequence marks 
the region of BSSs. \emph{Upper right panel:} assigning a Kurucz spectrum to the star F280. 
In such a way we can actually determine the parameters of the stars (lower-left panel). 
\emph{Lower right panel:} the ISEDs of the cluster with different combinations of spectral energy 
contribution to account for the stochastic effect. }\label{fig1-m67}
\end{figure}

BSSs are common and easily identified in colour-magnitude diagrams\index{colour-magnitude diagram} (CMDs) of star clusters. 
They are members of the host cluster and
located above and blueward of the cluster's main-sequence (MS) turn-off. BSSs in a cluster are obviously the 
most luminous blue stars in the CMD (see the top-left panel of Fig.~\ref{fig1-m67}), 
therefore they are the primary contributors in the total light of the cluster. The standard theory 
of single-star evolution cannot explain the presence of BSSs in SSPs' CMDs, 
and thus the standard SSP models do not include the contributions of BSSs. 
All currently accepted scenarios of BSS formation are related to stellar interactions. 
Coalescence in primordial binaries can launch BSSs to positions up to 2.5 magnitudes
brighter than the MS turn-off\index{turn-off}~\cite{McCrea64,CH09}.  
Mergers of binary-binary systems can produce possible BSSs with masses four times those of 
stars at the MS turn-off~ \cite{LL92}. Given the high luminosities and common 
presence of BSSs in stellar systems (e.g., \cite{AL95}
for open clusters\index{open cluster}; \cite{Piotto+02} for globular clusters\index{globular cluster}; 
\cite{Mapelli+09} for dwarf galaxies\index{dwarf galaxy}), 
we believe that we must consider the effects of BSSs in studies of stellar populations\index{stellar population} using 
population synthesis in unresolved observations. The key issue is how to accurately 
include BSS contributions in SSP models.

Studies show that no single mechanism can account for the entire BSS population observed in 
any one star cluster \cite{Stryker93}. This means that it is not easy to theoretically measure 
the respective contribution of BSSs in SSPs of different ages and metallicities. 
Although the detailed physical properties of BSSs can be better understood now 
(see Chap. 3, 8, 9, and 12), a comprehensive 
and robust model for BSS content in star clusters is still missing. Therefore, 
building up BSS population characteristics empirically from the statistics of a large sample 
of star clusters could be more practical and reliable than relying on incomplete theoretical approaches. 
In this way, the behaviour of BSSs (in terms of their specific frequency and relative distribution 
with respect to the MS turn-off in CMDs) can be modelled. Open clusters (OCs) in the Galaxy have a number of advantages 
for use as a working sample: \emph{(i)}most have photometric\index{photometry} data, thus enabling accurate determination 
of their fundamental parameters; \emph{(ii)} many have multi-epoch proper motion\index{proper motion} and/or radial velocity\index{radial velocity} data, 
so that their cluster membership probabilities can be measured accurately; and \emph{(iii)} they are 
representative of the environments of stellar populations shortly after their birth. 
Compared to OCs, globular clusters (GCs) are usually much more massive and often contain multiple stellar 
populations (e.g., \cite{Piotto08}), which renders them questionable as SSPs.

Preliminary modelling of BSS effects has been done on the basis of individual clusters in our 
previous papers \cite{Deng+99,XD05,Xin+07,Xin+08}, where we \emph{(i)} introduced 
the method used for calculating the realistic, integrated spectrum of a star cluster including the 
contribution of BSSs; \emph{(ii)} analysed the modifications to the integrated spectra and the broad-band colours 
caused by BSSs; and \emph{(iii)} estimated the possible uncertainties in the conventional SSP models, 
showing that the ages of star clusters\index{star cluster} can be underestimated by up to 50\% if BSS contributions are not considered.
To overcome such a great discrepancy,  we present a set of BSS-SSP models based on a statistical study of galactic OCs. 
Our models cover the wavelength range from 91 {\AA} to 160 $\mu$m, ages from 0.1 to 20~Gyr and 
metallicities $Z=0.0004, 0.004, 0.008, 0.02$ (solar metallicity\index{metallicity}) and 0.05.

This chapter is organised as the follows. In Sec.~\ref{M67}, 
we demonstrate how BSSs can affect the integrated spectral properties of a star cluster, 
taking one of the most studied old clusters, M67\index{M67}, as an example. In Sec.~\ref{OCs}, 
we present the statistical results of the properties of BSS populations based on 100 Galactic OCs, 
supplemented by a few rich GCs in Magellanic clouds\index{Magellanic Cloud} in Sec.~\ref{MCCs}.
In Sec.~\ref{BSSSP}, we describe the construction procedure of our models, 
as a generalised method based on that for M67, and we discuss
our model results and compare them with those of Bruzual \& Charlot (\cite{BC03}, hereafter BC03).  
Finally, a summary, brief discussion and prospectives  are presented in Sec.~\ref{discuss}.

\section{M67: Setting up the Scheme}\label{M67}\index{M67}
We started to consider quantitatively the effects of BSS on the integrated spectral 
properties of star clusters, and to constrain the theoretical models of SSPs in 1999 
using the well studied and relatively
rich galactic OC, M67\index{M67} \cite{Deng+99}. 

Almost all bright stars in the cluster region (about 1 degree in diameter) have reliable membership 
information, which is very important to the argument we want to pursue. 
This cluster were observed using a 15 intermediate-band filter photometric\index{photometry} system covering 
the whole optical wavelength range between 3000-10000~\AA\, (BATC survey, \cite{Fan+96}). Such an 
observation is ideally suited for our purpose. Figure~\ref{fig1-m67} demonstrated the key 
steps to build the integrated spectral energy distribution (ISED)\index{integrated spectral energy distribution} of the cluster:

\begin{itemize}
\item First of all, we need to find out the BSS members in the cluster. 
The upper left panel is a CMD\index{colour-magnitude diagram} of the cluster in two passbands 
(centred at 3890~{\AA} and 6075~{\AA}, respectively), with
the $x$-axis being the colour --- defined as the difference between magnitudes in two magnitudes. 
All the BSSs confined by the left box at the upper bluer extension of the MS
are almost sure members of M67\index{M67}. 
\item The upper right panel show how precisely we can anchor a \emph{Kurucz} spectrum to the 
observed spectral energy distribution\index{spectral energy distribution} (SED) determined in all the filters for the famous BSS member F280 in M67. 
In fact, a BSS star may not be a single star, therefore it is not possible to have parameters 
that are defined for single stars, such as effective temperature\index{effective temperature} and gravity\index{gravity}. 
Thus, the spectrum\index{spectroscopy} assigned to the star this way is a ``pseudo'' but still constitutes
an useful approximation. The spectrum assigned is indeed close enough and can represent the star's
spectral property. 
\item In this way, we can then accurately derive its ``physical parameters'', 
given the metallicity\index{metallicity} is solar as measured by numerous other observations (presumably also treated as single stars). 
This is demonstrated in the lower left panel. In our work, we assumed that the BSSs
can be taken as single stars in terms of spectral properties.
\item The ISED of the cluster is obtained by summing up individual SED of all bright members. To account for possible stochastic effects due to the 
low number of BSSs, we plotted different combinations of BSS sample in M67 (lower right panel).
\end{itemize}

As a conclusion, the ISED of M67\index{M67} built this way is apparently different from that of 
the theoretical SSP model corresponding to its age and metallicity. We further point out that, 
with the existence of BSSs, and when observed at unresolved conditions (such as in remote galaxies), 
the actual stellar population\index{stellar population} may appear substantially younger and/or more metal poor, 
sometimes less total mass compared to standard SSP models. M67 is just a typical old age cluster 
in our Galaxy\index{galaxy}, that can represent real stellar population of similar age and metallicity in any other galaxies,
which means, the widely used conventional SSP models in EPS need to be modified in order to be correctly 
applied to the analysis of stellar content in galaxies.

Thanks to the accurate but elaborate 15 intermediate band photometry\index{photometry} \cite{Deng+99}, the technique 
to build the ISED of M67 based on photometric observations is very unique and reliable. However, BATC 
photometry is not available for the entire galactic OC sample. Other photometry, including broadband, can also 
be helpful. The methodology is defined in Sec.~\ref{BSSSPLib}.

\section{ISEDs of Galactic Open Clusters}\label{OCs}\index{open cluster}
Building up a library of ISEDs of a large sample of star clusters\index{star cluster} 
is an observational approach towards realistic SSP models. Such a library can also be 
used to constrain theoretical efforts towards the same aim. By using the existing 
photometry and membership information of galactic star clusters, we have carried out 
a series of work in the past few years \cite{XD05,Xin+11,Xin+08,Xin+07}. 

Galactic OCs are the most studied stellar systems in the Galaxy, 
and have relatively more information about their physical properties including membership, 
age, metallicity, etc. 
However, the content of BSSs in OCs is rather stochastic due to the limited number 
of stars and BSSs, while all the BSS population may involve different ways of formation, 
and can be affected by dynamical processes in the cluster and with their environment.  

In \cite{XD05}, a small sample of 27 Galactic OCs from Ahumada \& Lapasset (\cite{AL95}; AL95 in the following)  
was selected to discuss the BSS contribution in a statistical way. 
The BSS population is better presented in the CMD of older OCs, 
thus the 27 OCs are all older than 1~Gyr, and provide
a constraint to SSP models of the same age and metallicity range.

The basic parameters of the selected clusters are given in
Table 1 of \cite{XD05}, where columns (1)--(3) give the cluster name, 
right ascension, and declination (J2000.0); columns (4)--(7) are the
ages, colour excesses [$E(B-V)$], distance moduli (DMs) and
metallicities ($Z$) of the clusters; the number of BSSs, $N_{\rm BSS}$, and
$N_2$ values are listed in columns (8) and (9), respectively; and finally, 
column (10) is the reference number. Values of $N_{\rm BSS}$ and
BSS photometric data for selected clusters are quoted directly
from AL95. The $N_2$, defined  
as the number of stars within an interval of 2 magnitudes below
the turn-off point for a given cluster (see detailed discussion in Sec.~\ref{BSSSP}), is also from AL95. The
other basic parameters of the selected clusters, specifically the
age, $Z$, $E(B-V)$, and DM, are extracted from the more recent photometric and theoretical work when results newer than
AL95 are available.

$N_2$ is regarded as a very important parameter that indicates
the richness of the theoretical stellar component in the simple
population synthesis scheme. For each cluster, this count is
made in the same CMD from which the BSSs are selected.
Then the ratio $N_{\rm BSS}$/$N_2$ can be taken as a specific BSS frequency
in a cluster and can be used as a probe for the cluster internal
dynamic processes concerning BSS formation. 

It is worth emphasising here that not all the clusters have
complete membership determinations in terms of both proper
motion\index{proper
motion} and radial velocity\index{radial velocity}. The selection of BSS candidates in
AL95 is basically according to where they appear in the observed CMDs. 
In our work, all the BSSs are treated equally,
without any further detailed membership measurements and
BSS identification. 

In order to demonstrate the effects of BSSs on the conventional ISEDs 
in the case of old OCs, we construct the ISED of the normal star 
and the BSS population in one cluster separately. The steps of 
how we build the composite SSP models including BSS contribution 
will be presented in details in Sec.~\ref{BSSSP}. The results with the OC samples were rather stochastic 
due to the small number of BSSs, or because of the small total number of 
stars in a given cluster. The ISEDs vary from BSS dominant --- one or a few very bright BSSs ---
or one in which BSSs contribute nothing to the ISED (see Fig.~4 in Xin, Deng \& Han~\cite{Xin+07}, 
and discussions on statistics there. This remind us that rich star clusters should be used, 
or find a proper way to enhance statistics of the OC sample such as binning the 
clusters in parameter space (see Section~\ref{BSinSP}).

\section{The Massive Star Clusters in the LMC \& SMC}\label{MCCs}
Star clusters in the Magellanic Clouds\index{Magellanic Cloud} are massive and have a large span in age. 
Unlike the globular clusters\index{globular cluster} in the Galaxy, they represent much better stellar populations 
in age coverage, and therefore serve as great targets for our work. In this section, 
the LMC cluster ESO~121--SC03 is analysed as the first example
to detect BSS contributions to the conventional SSP models using 
a massive intermediate-age cluster in a low-metallicity environment. ESO~121--SC03
is a distant northern Large Magellanic Cloud\index{Large Magellanic Cloud} (LMC) cluster, 
lying at a projected angular separation of 10 degrees from the LMC centre. 
It is described as a unique LMC cluster by Mackey, Payne \& Gilmore (\cite{MPG06}), 
because it is the only known cluster to lie in the LMC age gap. 
A significant number of previous studies (e.g., \cite{BA86,MHS86,MPG06}) 
claim an absolute age for ESO 121-SC03 in the range of 8--10~Gyr. Mateo et al. \cite{MHS86} 
obtained [Fe/H] = $-0.9\pm  0.2$ combined with a reddening of E($B - V$) = 0.03 mag. Mackey et al. \cite{MPG06} derive 
[Fe/H] = $-0.97\pm 0.01$ and E(V-I) = $0.04\pm 0.02$ for the cluster. They also mark a region in the 
CMD used to define BSS candidates in the cluster, but they do not study the BSS population in detail.

\begin{figure}
\sidecaption
\includegraphics[width=7.5cm]{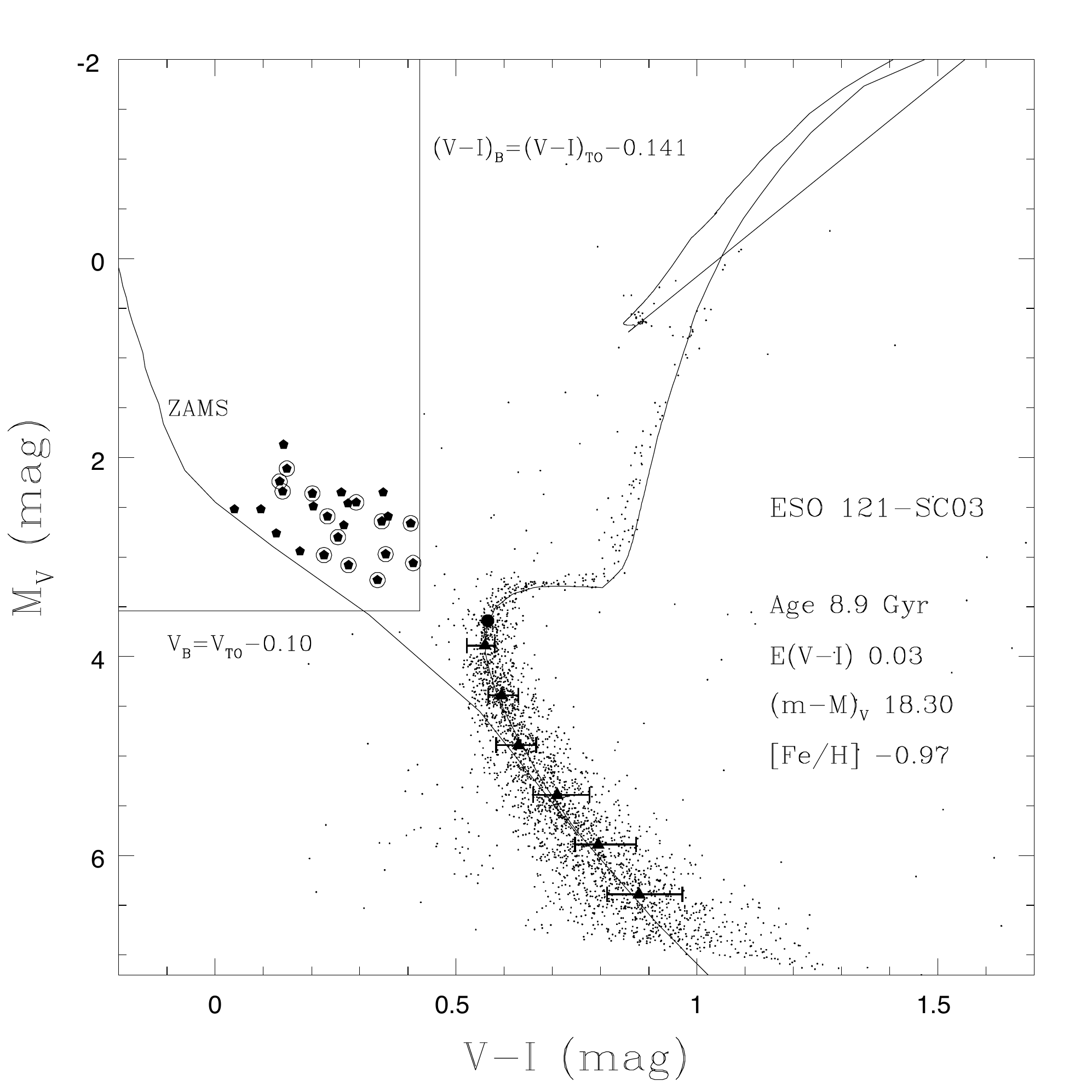}
\caption{Cleaned CMD\index{colour-magnitude diagram} for ESO~121-SC03\index{ESO~121--SC03}. 
The best-fitting Padova~2000 isochrone is overplotted.
The corresponding fit parameters are included in the figure legend.
In the CMD, we define a region that we will use for the
identification of the cluster's BSS population. The ZAMS is the
Padova~1994 isochrone at log(age yr$^{-1}$) = 6.60 and
[Fe/H]~$=-0.97$ dex. The solid bullet is the cluster's MS turn-off. 
The solid triangles represent the MS ridge line of the cluster. 
Pentagons are BSSs. Pentagons with circles are BSSs inside the half-light 
radius ($R_{\rm HL}$) of the cluster.}
\label{ESO121}
\end{figure}

\begin{figure}
\includegraphics[width=8cm]{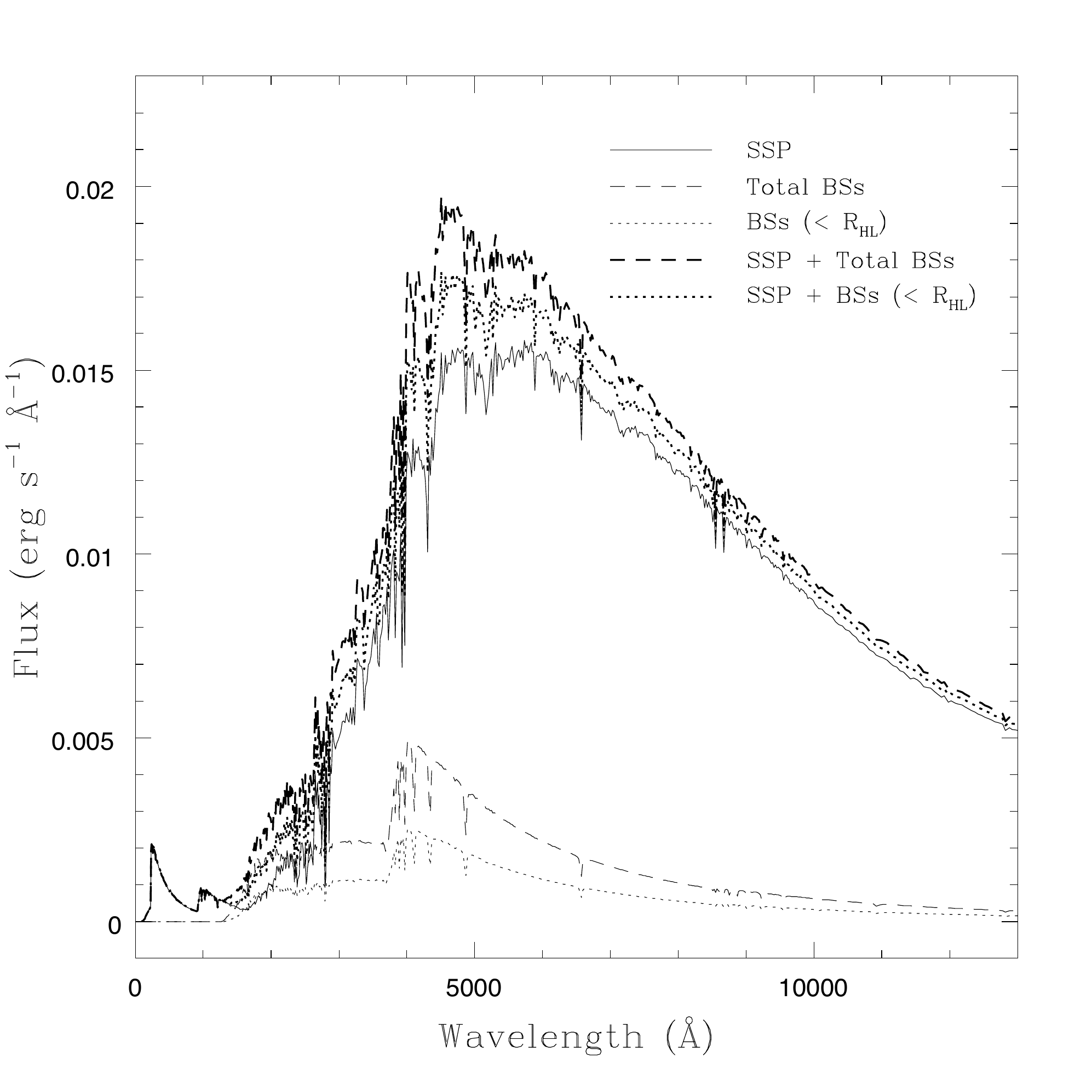}
\caption{ISED modifications. The BSS contributions are presented for two cases. 
The solid line is the ISED of the SSP component. The thin dotted line is the ISED of the BSS component 
for case (i): BSSs within $R_{\rm HL}$. The thin dashed line is the ISED of the BSS component for case (ii): 
all BSSs in the cluster. The heavy dotted line is the synthetic ISED of the SSP component and the 
case~(i) BSS component. The heavy dashed line is the synthetic ISED of the SSP component and the case~(ii) 
BSS component.}
\label{ESO121SED}
\end{figure}

As shown in Fig.~\ref{ESO121}, ESO~121-SC03
is a ``clean'' cluster and includes an obvious population of BSS candidates, 
which is very advantageous for our main aim of analysing the BSS contributions to 
the ISED of the cluster. This example demonstrated that, 
with statistics better than Galactic OCs, BSSs may have significant contributions to 
a stellar population \cite{Xin+08}. Figure~\ref{ESO121SED} shows the ISED of 
the cluster and how it is affected. With either BSSs only in the half-light radius ($R_{\rm HL}$)
or full sample in the cluster, apparent alternation due to BSSs to the theoretical ISED of its SSP 
model (the think solid line) is present. The change is stronger for the full sample, and the changes 
are always more prominent in the shorter wavelengths. This may actually infer either a younger age or 
lower metallicity when observed at unresolved conditions.

\begin{figure}
\includegraphics[width=10cm]{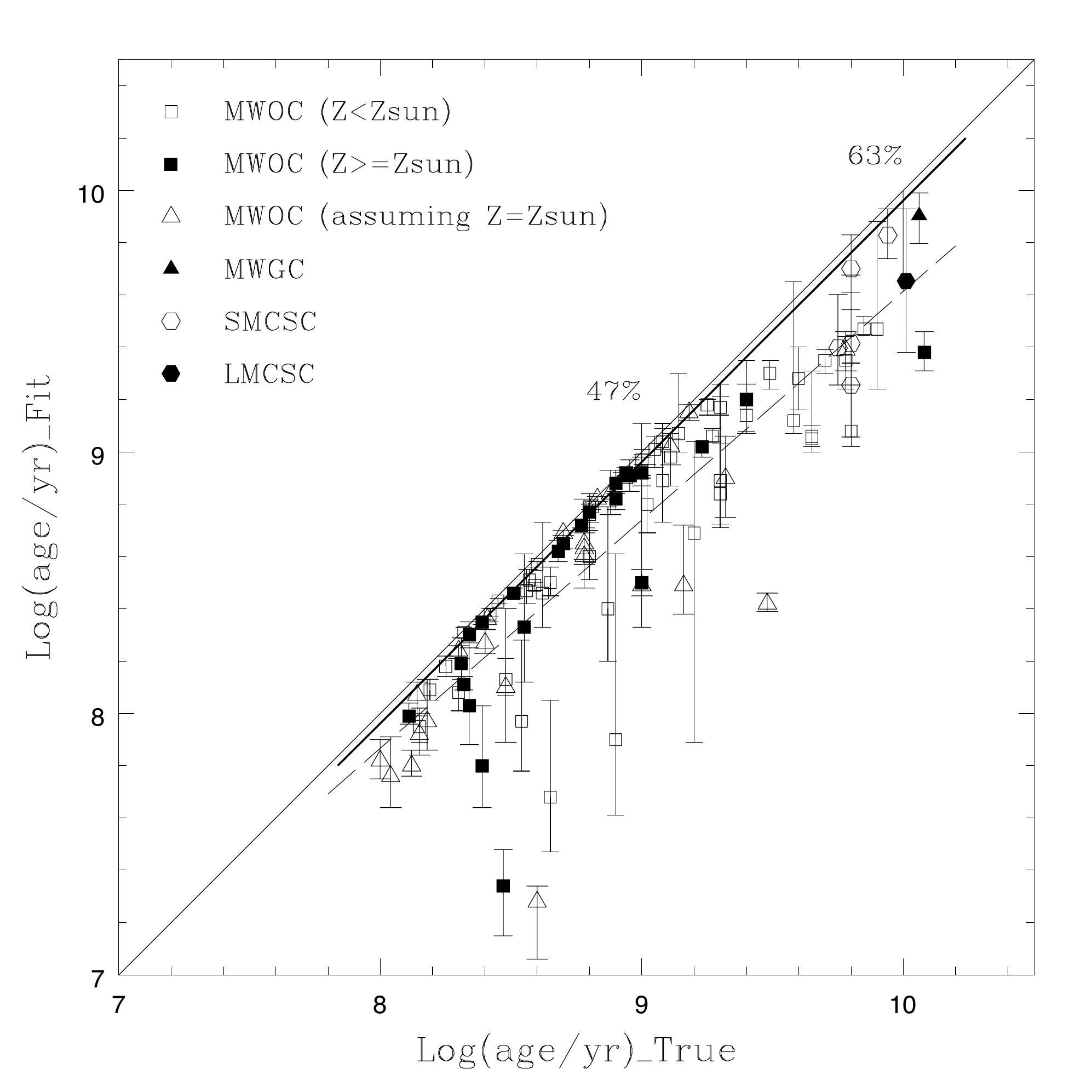}
\caption{The underestimation of age value by the conventional SSP models. 
The $x$-axis is the real age of star clusters, while the $y$-axis is the best-fitting age from the conventional SSP models. 
Different symbols mean different working samples. The ashed line is the least-square fitting of all the data 
points. The heavy solid line is given according to the formula of 
$t_b=(0.17+8.27Z_s)+(1.38-14.45Z_s)t_s$ \cite{LiHan08}, 
assuming 50\% binaries in stellar populations and using solar metallicity ($Z_s=0.02$). 
Distances between the diagonal and the dashed line are measured to detect the 
age uncertainty --- $(age_{real}-age_{fit})/age_{real}$. 
The underestimation of age value by the conventional SSP models becomes larger for older population. 
It is 47\% at 1~Gyr, and 63\% at 10~Gyr.}
\label{Agefit}
\end{figure}

Figure~\ref{Agefit} concludes our work on individual cluster's base. 
If we fit the ISED of star clusters\index{star cluster} by using the conventional SSP models, and we compare the ages\index{age} 
of the cluster measured by observations of clusters (taking as true age, the horizontal axis), 
to that of the best fit SSP model (the vertical axis), it is very clear that the fitted ages are 
substantially younger than the true ages of the clusters for the whole sample of clusters (all types). 
Quantitatively, at the true age of $\sim$1~Gyr, the deviation is 47\%, and is 63\% at the age of 
$\sim$10~Gyr. This fact tells us that EPS may have seriously underestimated ages for populations 
within the age range covered by our sample. 

Certainly, a larger sample of such rich clusters with good coverage in both age and metallicity\index{metallicity} 
is highly desired in order to constrain the method we proposed. Observational aspects, 
including detailed individual BSS, statistics of all BSSs in, and their dynamical effects on, 
rich GCs can be found in Chap. 5.

\section{Building Up an Empirical SSP Library}\label{BSSSP}

In order to create empirically a library for use in EPS, a dataset of observed star clusters covering 
wide enough parameter space (age, metallicity) has been made available in our previous work. 
In this section, we are going to describe how the library is actually built.

\subsection{A General BSS Distribution Function in Stellar Populations}\label{BSinSP}

Basically, two properties of BSS populations are relevant to the
modelling of BSS behaviour in SSPs, i.e. the number of BSSs ($N_{\rm BSS}$) 
in the SSP and their distribution in the SSP's CMD. In this
paper, both properties are obtained empirically from the observed OCs' CMDs. 
A working sample including 100 Galactic OCs\index{open cluster} from the catalogue
of AL95 is adopted to secure the reliability of the statistical results.

\begin{figure}
\includegraphics[width=10cm]{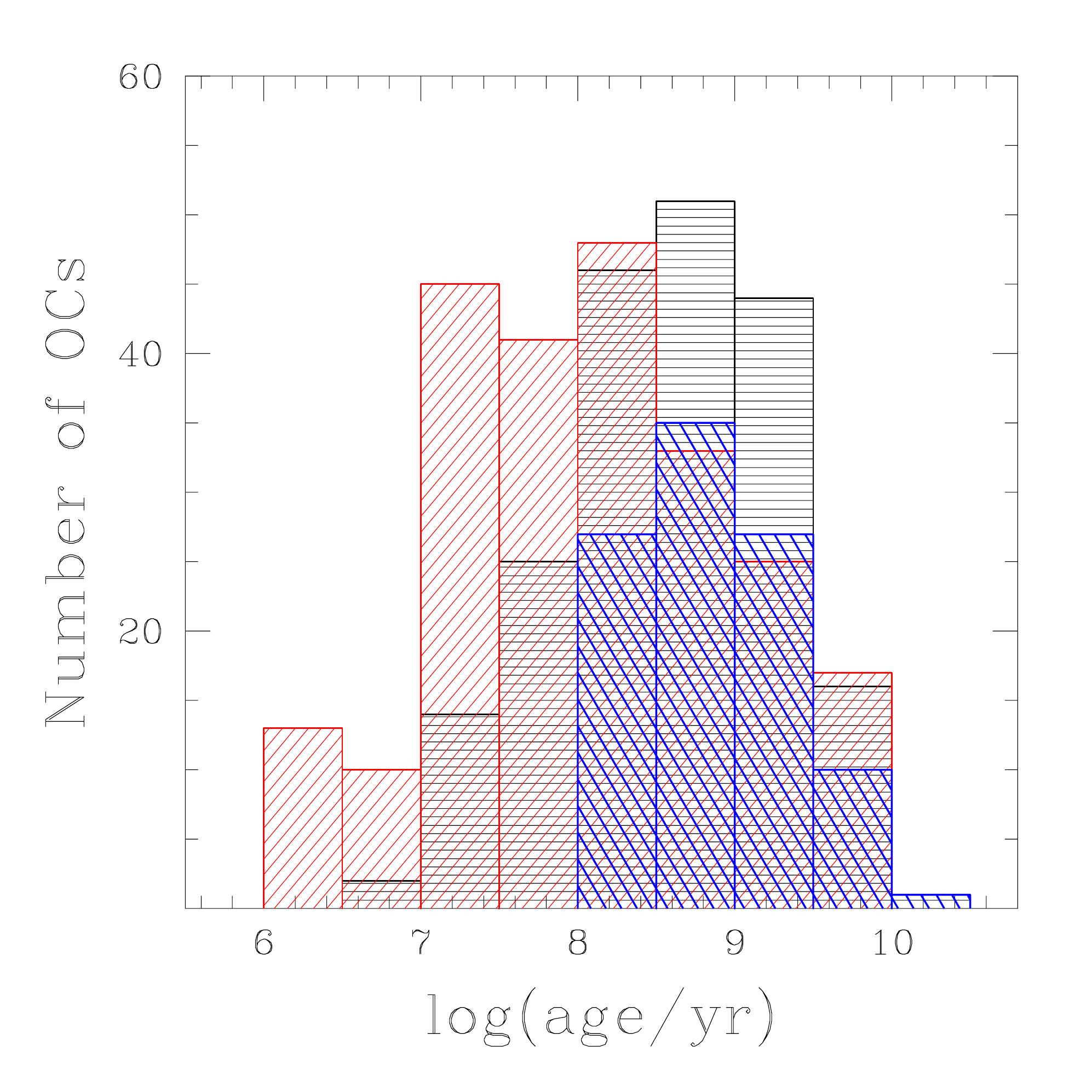}
\caption{Number distributions of Galactic OCs containing BSSs as a
  function of age. Horizontal shading represents the statistics from
  the Ahumada \& Lapasset \cite{AL07} catalogue. The hatching from the
  bottom left to the top right represents the statistics from AL95, while that from the top left
  to the bottom right shows the number distribution of our working
  sample (100 Galactic OCs).}
\label{AL07}
\end{figure}

Figure~\ref{AL07} shows the differences among the number distributions
of OCs containing BSSs versus age for the Ahumada \& Lapasset (\cite{AL07},
hereafter AL07) catalogue (horizontal hatching), the AL95 catalogue
(slanted hatching from bottom left to top right) and our sample
(slanted hatching from top left to bottom right). AL95 and AL07
published the most complete catalogues to date in terms of photometric
data of BSSs in Galactic OCs. They include almost all OCs containing
BSSs in the solar neighbourhood. Figure~\ref{AL07} shows that, compared
to AL95, AL07 dramatically reduced the number of young OCs containing
BSSs, mainly because of the difficulty of identifying BSSs in the CMD of
a star cluster that does not exhibit at least a fully developed
red giant branch (RGB) phase (particularly when information  on membership probability
is lacking). Moreover, most of the BSSs in young OCs (i.e.,
log(age/yr)$<$8.0 in this work) from both catalogues are located
close to the MS turn-off point in the CMD, which means that it is hard
to distinguish BSSs from MS stars. Such BSSs cannot effectively modify
the spectral intensity of a cluster. Therefore, we start the selection
of our working sample based on OCs with log(age/yr)$\geq$8.0.

Meanwhile, to construct a spectrum of the BSS population for a given
OC, we need to obtain the physical parameters of the BSSs in the CMD of
the OC. To do this, we need the age\index{age},
metallicity\index{metallicity}, colour excess\index{colour excess} and distance\index{distance} modulus for each OC, which are
not all included in either AL95 or AL07. Therefore, we decided to keep
only the photometric data of BSSs from AL95, to ensure a homogeneous
selection of the BSS sample. We collected the remaining OC parameters
from the recent literature (see Table~1 in \cite{Xin+07}). In
practice, OCs with log(age/yr)$\geq$8.0 from AL95 are included as
sample clusters if reliable parameters can be found, in the sense that
most of the BSSs are located at reasonable positions in the CMD with
respect to the Padova~1994 isochrone for the OC's age and metallicity.

There are 33 OCs with age$\ge$1.0~Gyr in the combined catalogue
comprised of AL95, AL07 and our working sample. Older OCs have better
statistics as regards their BSS populations. Using the photometric data
of the BSSs in these 33 OCs as an example, we analysed the BSS distribution
functions versus the MS turn-off\index{turn-off} point, as a function of $M_V$ and $(B-V)$.
In such a way, we can detect differences in the BSS distributions in the CMD
between the AL95 and AL07 statistics. It is shown that the distribution functions 
of both $M_V$ and $(B-V)$
do not exhibit any essential differences between the AL95 and the AL07
catalogues. Therefore, we continue on the basis of the results from
our previous work, i.e., the parameters of the sample clusters and the
BSS population from AL95.

\begin{figure}
\includegraphics[width=12cm]{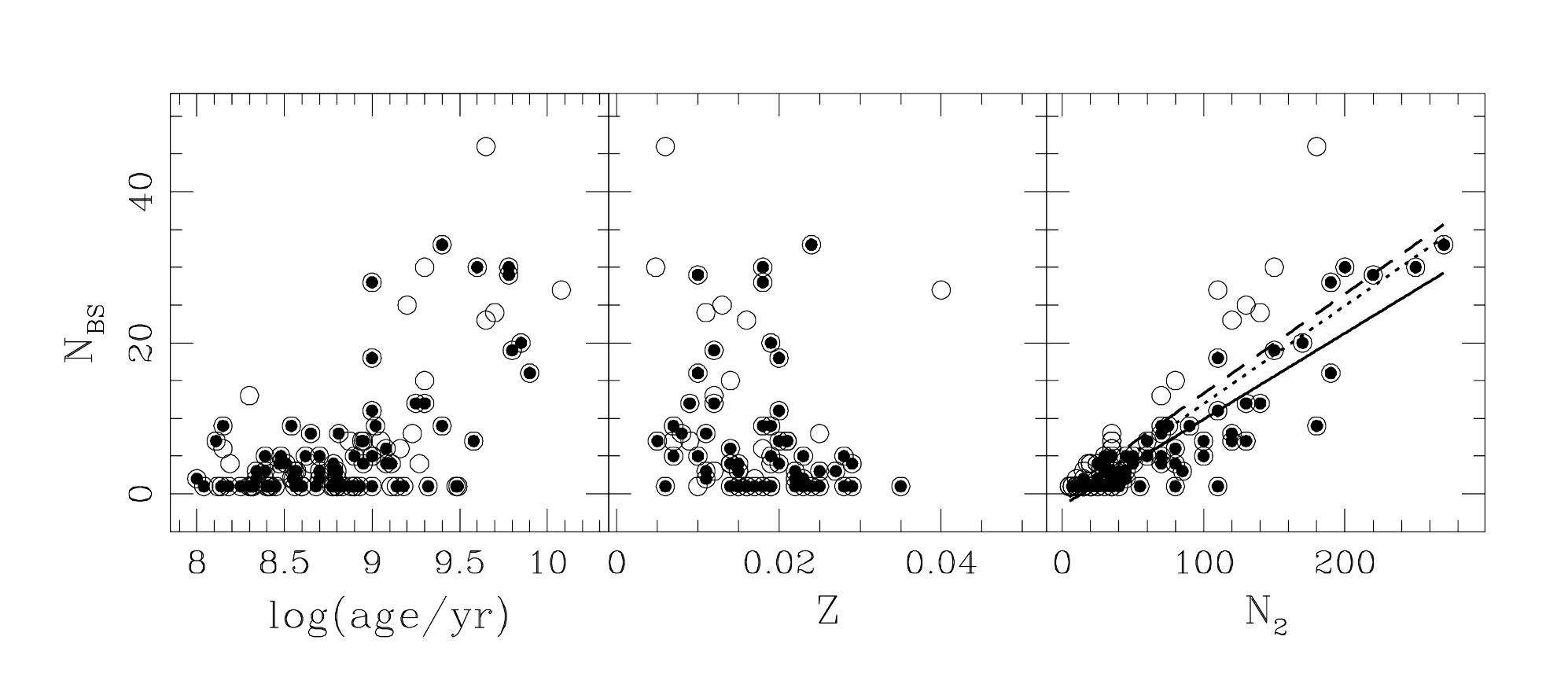}
\caption{Correlations (if any) between $N_{\rm BSS}$ and the age,
  metallicity and richness ($N_2$) of Galactic OCs, respectively. The
  open circles show the results for the 100 OCs in our working
  sample. The filled circles mark the OCs with $N_{\rm BSS}/N_2$ less
  than the average ratio plus $1\sigma$ (standard deviation). The
  solid line in the right-hand panel is the least-squares fit to the
  filled circles; it is given by Eq.~(\ref{eq1}). For reference, the
  dotted line in the right-hand panel is the least-squares fit to all
  OCs (open circles), and the dashed line is the fit to OCs with
  ages$\geq$1.0 Gyr.}
\label{fig3}
\end{figure}

\begin{figure}
\sidecaption
\includegraphics[width=7.8cm]{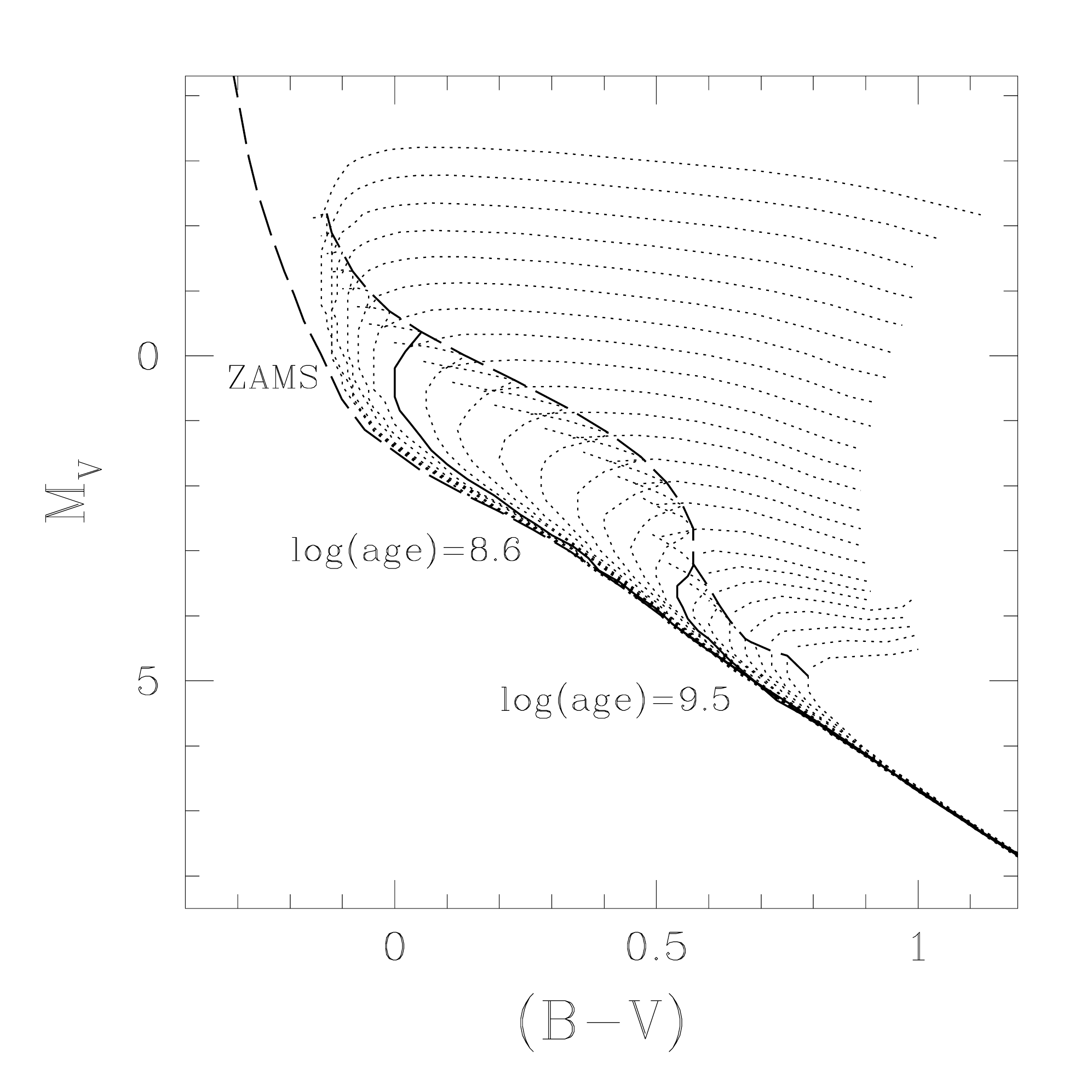}
\caption{Criteria used for our age-bin selection. The dotted lines are
  the isochrones\index{isochrone} for ages between 0.1 and 20 Gyr, truncated at the
  bottom of the RGB phase. The two dashed lines mark the
  MS stage between the zero-age MS (ZAMS) and terminal-age MS (TAMS). 
The choice of age bin is shown as the solid lines.}
\label{fig4}
\end{figure}

\begin{figure}
\includegraphics[width=5.5cm]{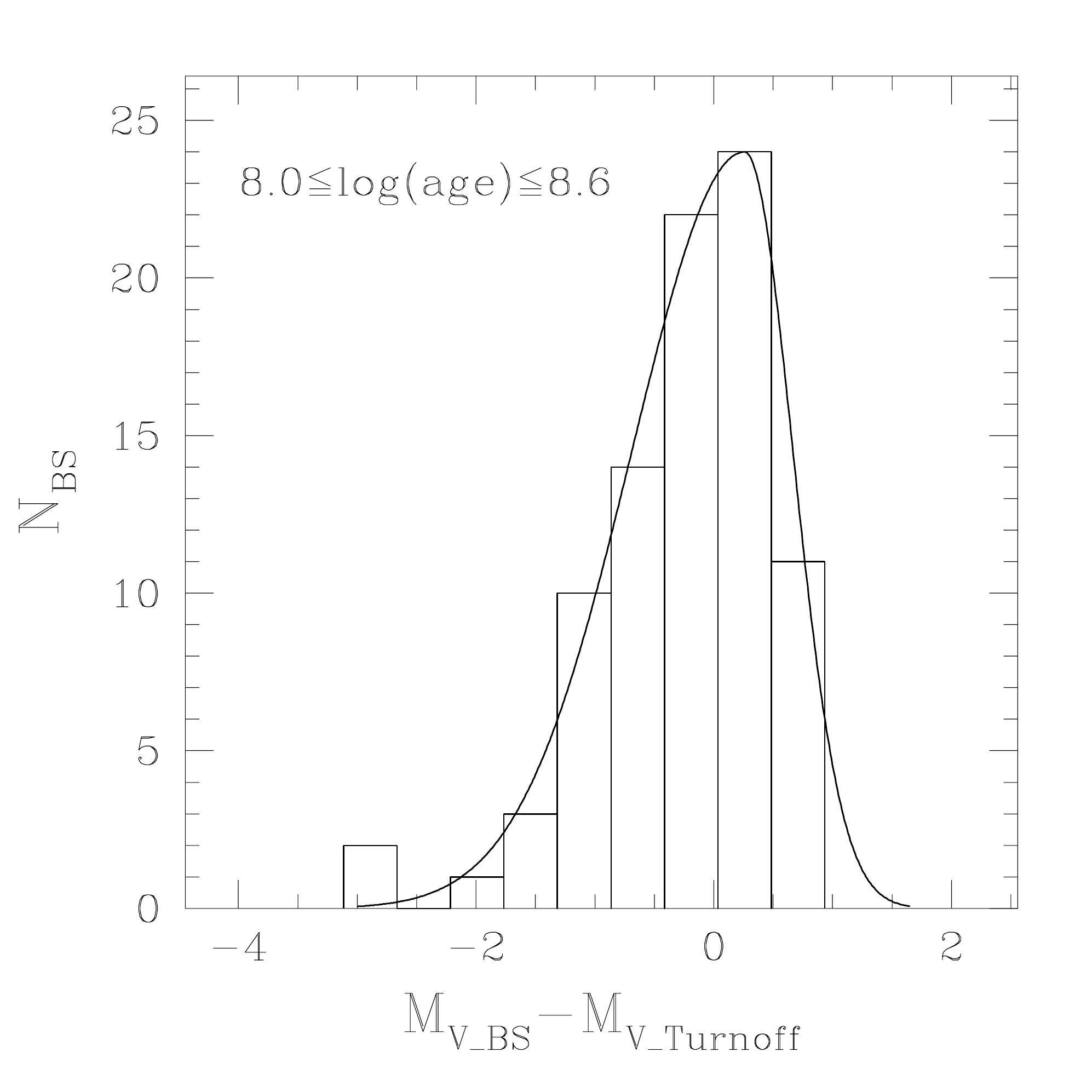}\includegraphics[width=5.5cm]{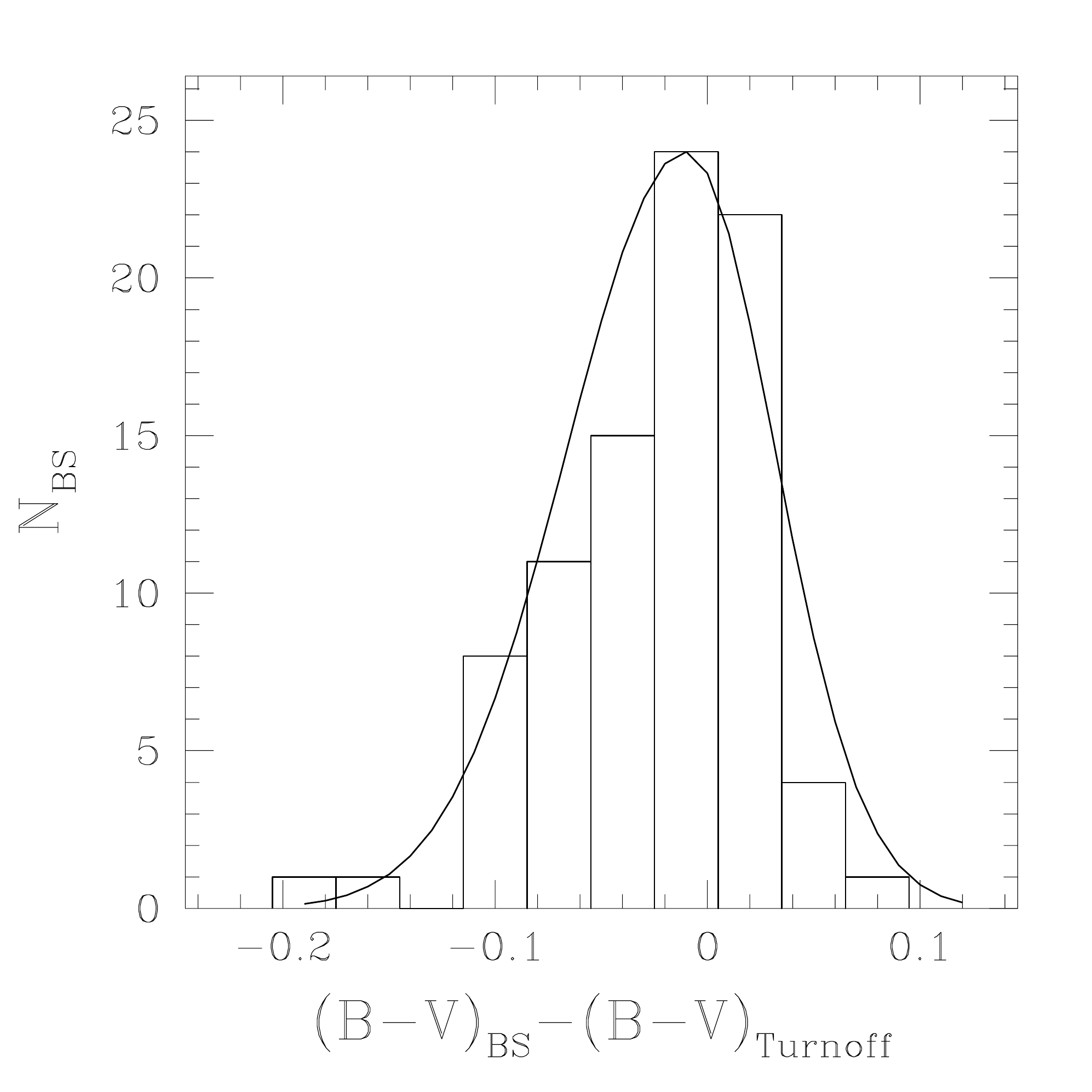}\\
\includegraphics[width=5.5cm]{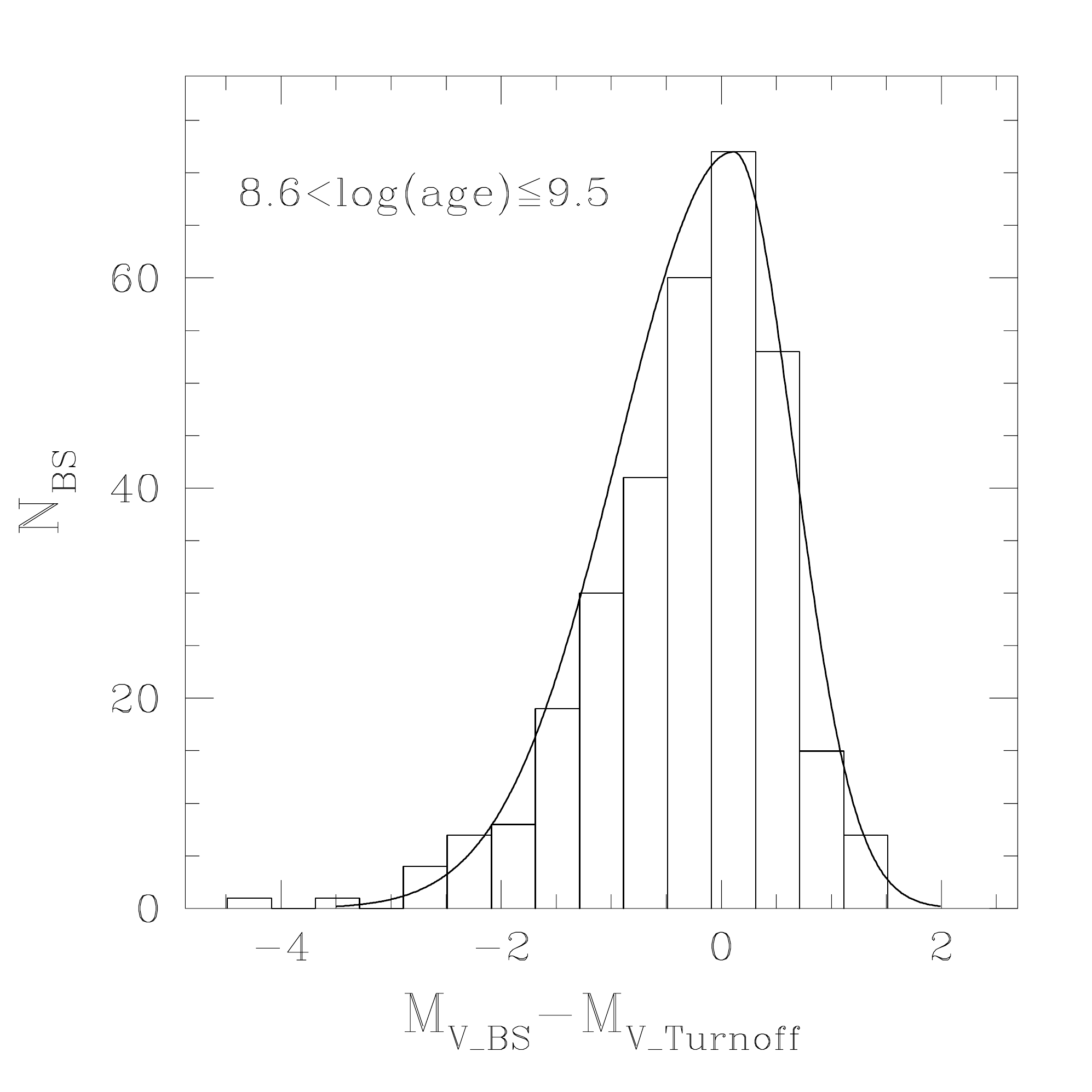}\includegraphics[width=5.5cm]{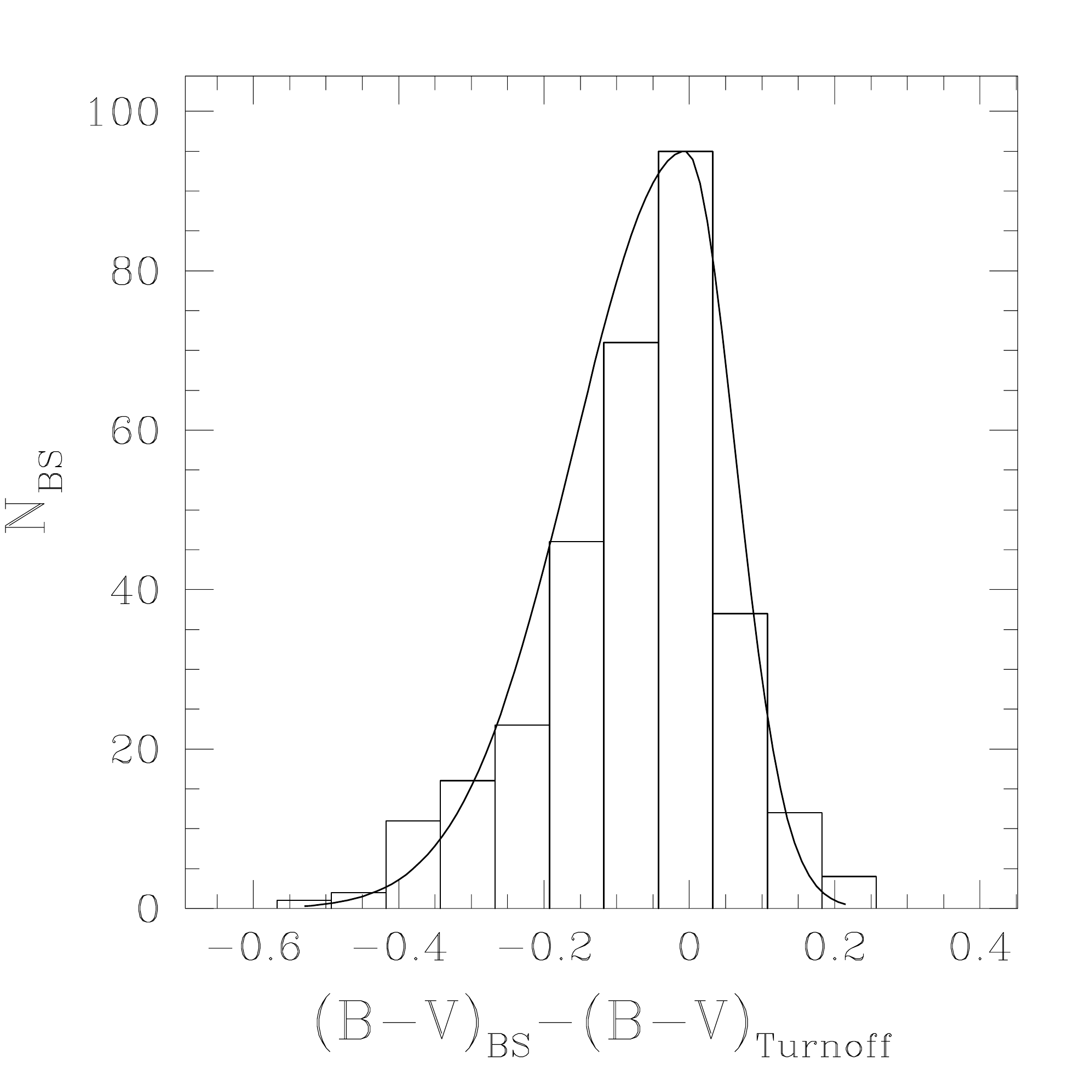}\\
\includegraphics[width=5.5cm]{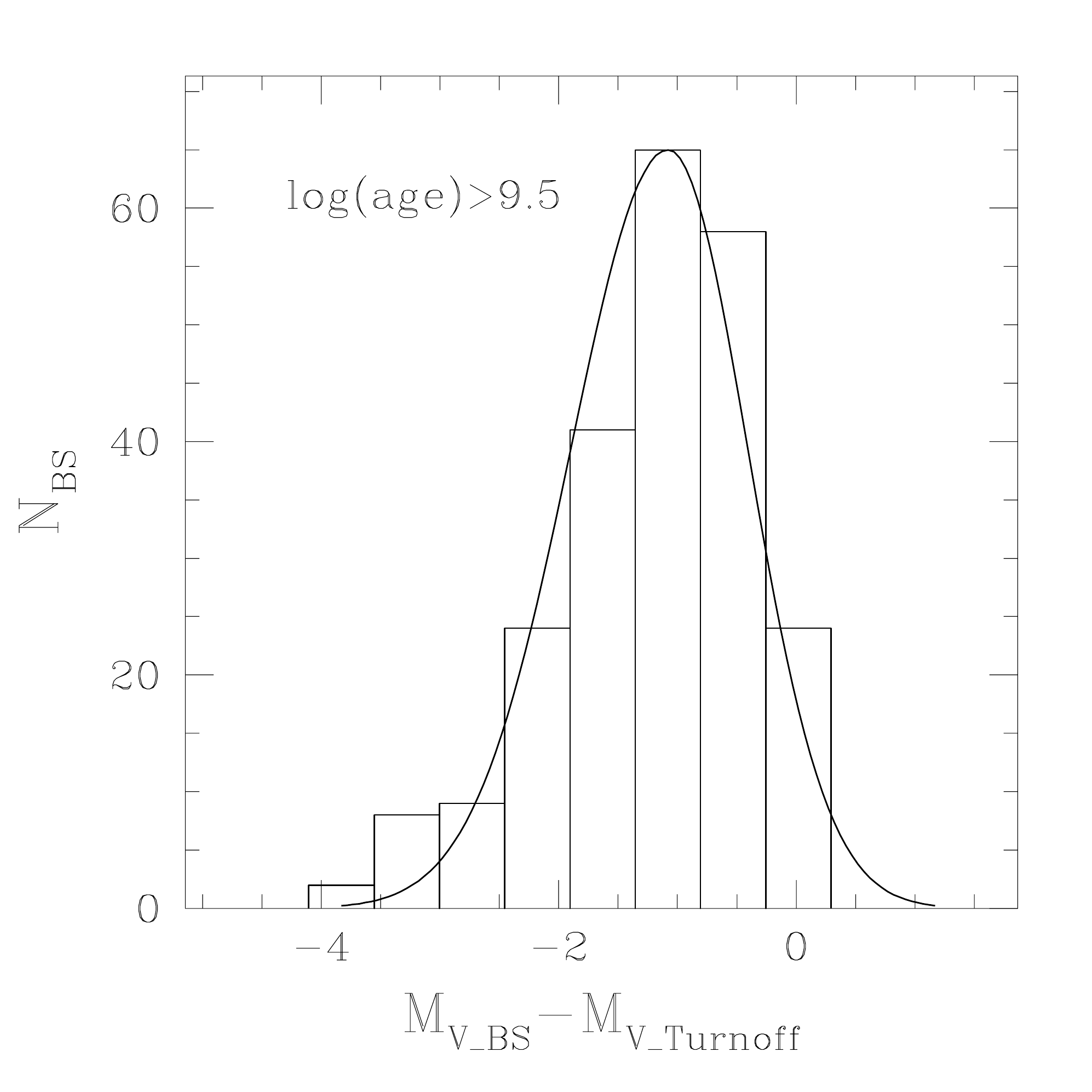}\includegraphics[width=5.5cm]{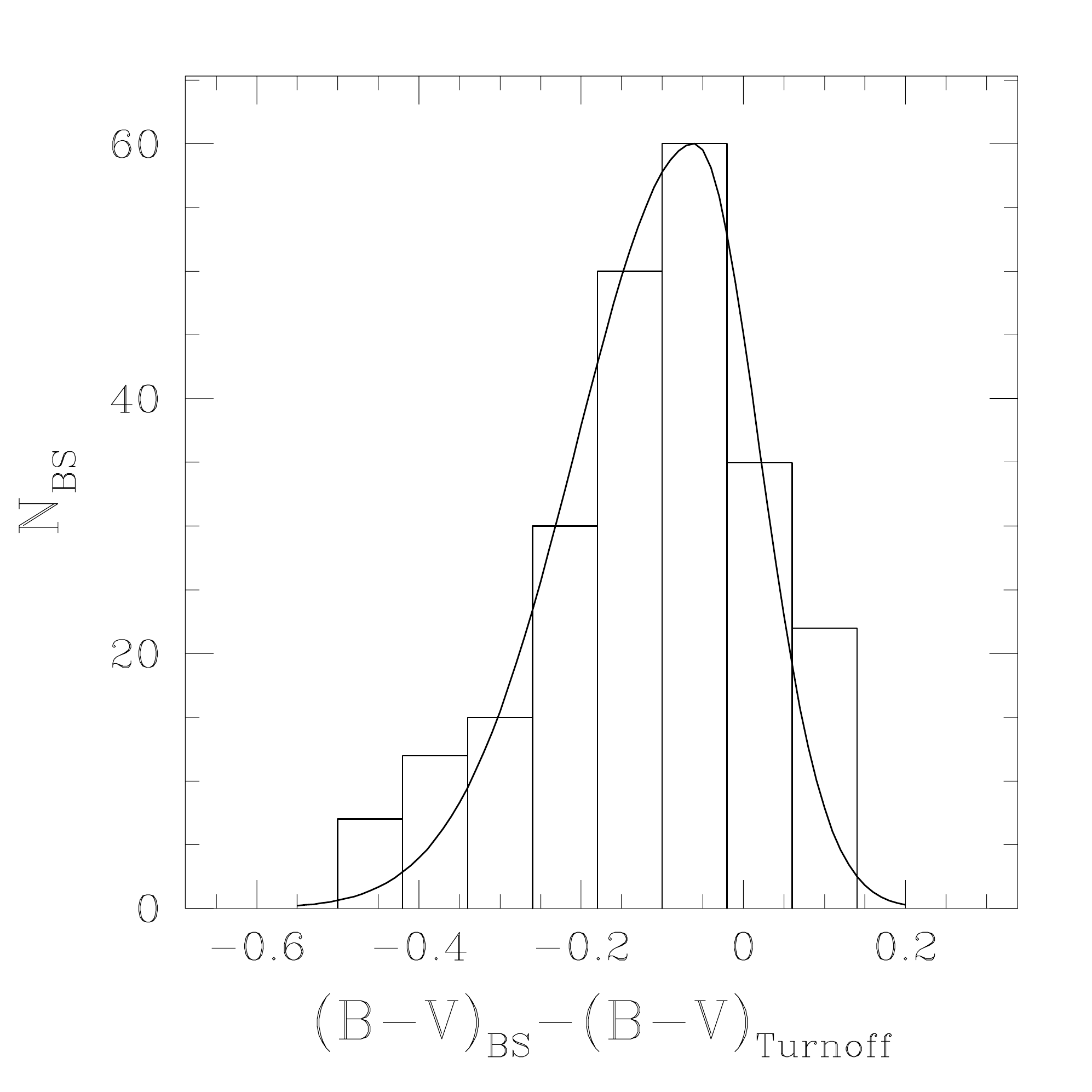}
\caption{BSS distribution functions in our OCs' CMDs in three different
  age bins as a function of $M_V$ (left panels) and the $(B-V)$ (right
  panels). In each panel, a gaussian profile is used to describe
  either side of the peak separately.}
\label{fig5}
\end{figure}

Details of the BSS properties are presented in
Figs.~\ref{fig3}--\ref{fig5}. Figure~\ref{fig3} shows $N_{\rm BSS}$ as a
function of the age, metallicity and richness of the sample OCs. The
richness of a star cluster is represented by $N_2$, which is the
number of cluster member stars within 2 magnitudes below the cluster's
MS turn-off. The open circles represent the results for the 100
Galactic OCs. Because of the small number of member stars and even
smaller number of BSSs in the individual OCs, the results directly
derived from Fig.~\ref{fig3} are very stochastic, and thus we use the
ratio of $N_{\rm BSS}$/$N_2$ to reduce the effects of stochasticity. We
use this ratio as definition of the specific frequency of BSS
components in SSPs. We calculated the standard deviation ($\sigma$) of
the ratio for the entire sample, i.e., $\sigma=\sqrt
{\frac{\sum_{i=1}^N(\frac{N_{\rm BSS}}{N_2}_i-\overline{\frac{N_{\rm
BSS}}{N_2}}~)^2}{N\times(N-1)}}$ and $N=100$. We marked the
OCs with $\frac{N_{\rm BSS}}{N_2}\le \overline{\frac{N_{\rm BSS}}{N_2}}
+ 1\sigma$ with filled circles in Fig.~\ref{fig3}.

The left-hand panel in Fig.~\ref{fig3} shows $N_{\rm BSS}$ versus
cluster age on a logarithmic scale. $N_{\rm BSS}$ seems largely
insensitive to age until a sudden increase for ages greater than
1.0~Gyr. Based on this figure it is hard to discern any correlation
between the two parameters, and it is also risky to jump to the
conclusion that $N_{\rm BSS}$ is not correlated with age. The seemingly
constant $N_{\rm BSS}$ for age$<$1.0~Gyr could be caused by the
confusion of defining an accurate MS turn-off point and a BSS population
in relatively young star clusters. In our model construction, we have
not adopted any correlation between $N_{\rm BSS}$ and age. The only
correlation we used is that between $N_{\rm BSS}$ and $N_2$ (shown in
the right-hand panel in Fig.~\ref{fig3}; see below). In fact, for
SSPs, $N_{\rm BSS}$ and age are related through the SSP's $N_2$. $N_2$
increases following the initial mass function (IMF)
slope as the SSP ages, and so does $N_{\rm BSS}$ 
through the correlation between $N_{\rm BSS}$ and $N_2$.

The middle panel of Fig.~\ref{fig3} shows $N_{\rm BSS}$ as a function
of metallicity for our sample OCs with published metallicity
information. No correlation can be established. We previously studied
whether the BSS strengths are sensitive to metallicity. We did not find
any obvious correlation between these two parameters either (see Fig. 18 in \cite{XD05}).

The right-hand panel of Fig.~\ref{fig3} clearly shows that only
$N_{\rm BSS}$ and $N_2$ are correlated. The implication of this
correlation is that $N_{\rm BSS}$ is proportional to the richness of an
SSP, but not (at least not obviously) to any other parameter, such as
age or metallicity. The solid line in the right-hand panel is the
least-squares fit to the filled circles. For reference, we present two
more fits for different samples. The dotted line is the least-squares
fit to all OCs (open circles), while the dashed line is the fit to
those OCs that are older than 1.0~Gyr. It is hard to tell which fit is
most accurate. We choose the solid line for estimating $N_{\rm BSS}$ in
an SSP simply to avoid exaggeration of the BSS-enhanced intensity in
our SSPs. The correlation can be empirically described as
\begin{equation}
N_{\rm BSS}= (0.114\pm0.006) \times N_2 - (1.549\pm0.731). \label{eq1} 
\end{equation}
The uncertainties in the coefficients result from the $1\sigma$
uncertainty in $N_{\rm BSS}/N_2$.

The loci of BSSs in cluster CMDs are fixed by their formation and
evolutionary processes \cite{Ferraro+09}, which could still be
stochastic owing to the varying physical conditions among star
clusters. Thus, it is impossible to construct the specific
distribution function of the BSS population for each individual SSP. A
feasible approach to generate a BSS population in the CMD is to work
out a uniform BSS distribution function.

The dotted lines in Fig.~\ref{fig4} are isochrones with ages between
0.1 and 20 Gyr, truncated at the bottom of the RGB phase. The two
dashed lines, i.e. the zero age MS (ZAMS)\index{ZAMS} and the boundary between
the MS and post-MS phases, highlight the entire MS stage for all
isochrones. The distance between the two dashed lines becomes narrower
in the colour range in the CMD\index{colour-magnitude diagram} for older -- log(age/yr)$\ge$9.5 -- and
younger -- log(age/yr)$\le$8.6 -- SSPs, which means that a uniform BSS
distribution function will certainly push a BSS population towards red
colours for SSPs with a MS turn-off close to the ZAMS. To avoid
uncertainties associated with the distance between the MS turn-off\index{turn-off}
points and the ZAMS, we consider BSS distribution functions in three
different age bins, i.e., 8.0$\le$log(age/yr)$\le$8.6,
8.6$<$log(age/yr)$\le$9.5, and log(age/yr)$>$9.5. The solid lines in
the figure show the age-bin selection.

Since the bin selection is done empirically, we tested that a small
change in the age-bin selection, i.e., $\Delta \log({\rm
  age/yr})=0.1$, cannot effectively modify the BSS distribution in the
CMDs and, consequently, affect our model results. Specifically, we
construct the BSS distribution functions with $\log({\rm age/yr})=8.6$
included in the older age bin and calculate the $(U-B)$ and $(B-V)$
colours of the SSP with $Z=0.02$ and $\log({\rm age/yr})=8.6$. The
colour changes resulting from adopting different distribution
functions are $<~0.0001$~mag for both colours. The same result is
found for a test with $\log({\rm age/yr})=9.5$.

We align the MS turn-off points of all OCs in our working sample to
obtain a sufficient number of BSSs for good statistics.
Figure~\ref{fig5} shows the distribution functions of BSSs versus the MS
turn-off point in three different age bins as a function of $M_V$ (left
panels) and $(B-V)$ (right panels). For each of the distributions, a
gaussian profile is adopted to fit either side of the peak separately,
based on which we construct the BSS population in the CMD of an SSP
using Monte Carlo simulations\index{Monte-Carlo simulation} in two dimensions ($M_V$ and
$(B-V)$). All gaussian profiles have the same standard format,
\begin{equation}
f(x)=\exp \left[-\frac{\left(x-\mu\right)^2}{2\sigma^2}\right], \label{eq2}
\end{equation}
where $\mu$ is the peak position and $\sigma$ the standard deviation.
The notation $x=\Delta M_V=M_{V_{\rm BSS}}-M_{V_{\rm TO}}$ refers to
the increment in luminosity ($M_V$) of a BSS with respect to the SSP's
turn-off luminosity; $x=\Delta (B-V)=(B-V)_{\rm BSS}-(B-V)_{\rm TO}$ is
the equivalent increment in colour index $(B-V)$. The specific values
of $\mu$ and $\sigma$ for different $x$ ranges are listed in
Table~\ref{table2}. Here, $f^-(x)$ and $f^+(x)$ refer to the left- and
right-hand sides of the gaussian profiles, respectively.

Since the distribution functions are constructed empirically from
observational statistics, the values of both $\mu$ and $\sigma$ are
inevitably sensitive to the selection of the bin size. To find
reasonable descriptions for the gaussian profiles, we start exploring
the statistics of $M_V$ and $(B-V)$ in each age bin with a bin size of
$\Delta M_V=0.20$ and $\Delta (B-V)=0.010$ mag, and we subsequently
increase the bin size in steps of $\Delta M_V=0.05$ and $\Delta
(B-V)=0.005$ mag until the distribution resembles a gaussian
function. The adopted bin sizes for $\Delta M_V$ and $\Delta (B-V)$
for each age bin are given in column (3) of Table~\ref{table2}. The
$\mu$ and $\sigma$ values listed in the table are calculated based on
the corresponding bin size.

\begin{table*}
\caption{Gaussian profile parameters for BSS distribution functions in
  different age bins}
\label{table2}
\begin{tabular}{ccrrcccc}
\hline\noalign{\smallskip}
Age bin& &  bin size&$\mu$& \multicolumn{2}{c}{$f^-(x)$}&\multicolumn{2}{c}{$f^+(x)$} \\
(yr) & & (mag) & (mag) & $\sigma$  & $x$ range & $\sigma$ & $x$ range\\
\noalign{\smallskip}\svhline\noalign{\smallskip}
8.0$\leq$log(age)$\leq$8.6&$M_V$&0.45& 0.26 & 0.9454&[$-$3.00, 0.26]&0.4056&[0.26, 1.65]\\
                                      &$(B-V)$ &0.03 & $-$0.01 &0.0563 &[$-$0.19, $-$0.01] & 0.0418 & [$-$0.01, 0.12]\\
8.6$<$log(age)$\leq$9.5 &$M_V$&0.40&0.112 & 1.0464 &[$-$3.50, 0.112]&0.5470 &[0.112, 2.00]\\
                                      &$(B-V)$ &0.075&$-$0.005 &0.1544 &[$-$0.530, $-$0.005]&0.0678 &[$-$0.005, 0.220]\\
log(age)$>$9.5&$M_V$&0.55&$-$1.08 & 0.8155 & [$-$3.83, $-$1.08]&0.6711 &[$-$1.08, 1.20]\\
                             &$(B-V)$&0.08 & $-$0.06 & 0.1472 &[$-$0.55, $-$0.06]&0.0837 &[$-$0.06, 0.20]\\
\noalign{\smallskip}\hline\noalign{\smallskip}
\end{tabular}
\end{table*}

In addition to the distribution function, a further boundary that also
constrains BSS positions in the CMD is the ZAMS. Any BSSs located
beyond the ZAMS, i.e., with bluer colours than the ZAMS for the same
luminosity, will not be generated by our program. This assumption is
made mainly because we treat BSSs as MS stars and describe them using
standard MS models.

Our sample of 100 Galactic OCs has limited parameter coverage in age
and metallicity, i.e., it covers ages from 0.1 to 12~Gyr and
metallicities from $Z=0.0048$ to 0.035 (Xin et al. 2007, their table
1). To consider the BSS contributions for the full set of SSP models,
some extrapolations of the parameter space have been adopted based on
the results from OCs and some reliable working assumptions: \emph{(i)} ages
from 0.1 to 20~Gyr, with the lower limit set by the difficulty to
identify BSS components in star clusters younger than 100 Myr, while
the BSS properties are not expected to change dramatically in very old
stellar populations; and \emph{(ii)} metallicities $Z=0.0004,
0.004, 0.008, 0.02$ and 0.05. We extend the metallicity to a lower
boundary of $Z=0.0004$ and a maximum value of $Z=0.05$ because the
statistics of OCs show that BSS behaviour is not sensitive to
metallicity\index{metallicity}. ($Z=0.0001$ is not included in the models because
horizontal branch stars\index{horizontal branch star}, instead of BSSs, dominate the energy in the
ultraviolet\index{ultraviolet} and blue bands in extremely metal-poor populations.)

\begin{table*}
\caption{Fundamental ingredients of the BSS-SSP models}\label{table1}
\begin{tabular}{lll}
\hline\noalign{\smallskip}
Name & Property & Source\\
\noalign{\smallskip}\svhline\noalign{\smallskip}
Padova1994 isochrones & $Z$=0.0001--0.05~~Age=4Myr--20Gyr & Bertelli et al. (1994)\\
BaSeL spectral library  & 91{\AA}--160$\mu$m & Lejeune et al. (1997)\\
                                        & median resolution $\lambda/\Delta\lambda\approx300$ &\\
Initial mass function & $\xi(\log~m)\propto m^{-1.35}$ & Salpeter IMF (Salpeter 1955)  \\  
                                     & table~1 in Chabrier (2003)        & Canonical IMF (Kroupa 2002)\\
\noalign{\smallskip}\hline\noalign{\smallskip}
\end{tabular}
\end{table*}

\subsection{Building the Empirical SSP Library}\label{BSSSPLib}

With the distribution functions of BSS in a population, it is now possible to build an empirical library of stellar populations. The fundamental ingredients of our models are listed in
Table~\ref{table1}. For convenience, we use the widely adopted BC03
models as reference. Modifications owing to BSSs are calculated as
increments to the BC03 SSPs of the same age and metallicity. We
adopted the Padova~1994 isochrones\index{isochrone} \cite{Bertelli+94} and the {\tt BaSeL}
spectral library \cite{Lejeune+97} because they homogeneously
cover the widest ranges of age and metallicity, and the longest
wavelength range. The Padova~2000 isochrones \cite{Girardi+00} are
based on a more recent equation of state and low-temperature opacities
compared to Padova~1994. However, we decided against adopting them for
our model construction, because BC03 do not recommend to use their
SSPs based on the Padova~2000 isochrones.  They state that their models
based on the Padova~2000 isochrones 
\begin{quotation}
tend to produce worse agreement
with observed galaxy colours.
\end{quotation}

High-resolution observational spectral libraries (e.g.,
\cite{Pickles98,STELIB}) suffer from problems
related to limited parameter coverage. Instead of combining spectra
from different libraries to enlarge our parameter coverage, we decided
to use only the theoretical library.

For consistency with BC03, we adopted the Salpeter~\cite{Salpeter55}
and Chabrier~\cite{Chabrier03}
IMFs.
As clearly shown by \cite{Dabringhausen+08}\footnote{see their Fig. 8},
the Chabrier IMF is almost indistinguishable from the Kroupa IMF
\cite{Kroupa01,Kroupa02} when normalised as 
$\int_{0.1}^{100}\xi(m){\rm d}m=1{\rm M}_{\odot}$, 
which is exactly how both BC03 and we
ourselves normalise the SSP models. Using such a normalisation, the
slight differences between the two IMFs cannot cause any effective
modifications as regards the BSS contribution to SSPs. Therefore, we
refer to the IMF (in addition to the Salpeter IMF) as the ``Canonical
IMF'' throughout this work. It can be conveniently described by a
two-part power law, $\xi(m) \propto m^{-\alpha_i}$, with
$\alpha_1=1.3$ for the stellar-mass range $0.08 \le m/{\rm M}_\odot <
0.5$ and $\alpha_2=2.3$ for $m \ge 0.5 {\rm M}_{\odot}$ \cite{Kroupa01},
or in terms of a power law plus a lognormal form as presented in
Table~1 of \cite{Chabrier03}.

Our construction procedure for SSP models including BSS contributions
is summarised as follows: 
\emph{(i)} We use the standard models (we use BC03
in this work, but we can in principle use any other SSP flavour as
well) to represent the integrated spectrum of the ``normal'' SSP member
stars. We subsequently use the statistical properties of BSSs from
Galactic OCs to generate the BSS population for the appropriate
SSP; and 
\emph{(ii)} We calculate the spectrum of the BSS population and combine
it with the spectrum of the normal member stars after the appropriate
flux calibration. The composite spectrum is the integrated spectrum of
the BSS-corrected SSP.

In detail, the model construction includes the following steps:

{\bf 1)} We assume that any given model SSP contains $10^5$ original member
stars. The corresponding normalisation constant, $A$, for a given
IMF is calculated as
\begin{equation}
10^5 = A \times \int_{m_{\rm l}}^{m_{\rm u}} \phi(m)~{\rm
  d}m, \label{eq3}
\end{equation}
where $\phi(m)$ is the IMF, $m_{\rm l}=0.1 {\rm M}_{\odot}$ and
$m_{\rm u}=100 {\rm M}_{\odot}$.

{\bf 2)} The SSP's $N_2$ number is calculated using
\begin{equation}
N_2 = A \times \int_{m_1}^{m_2} \phi(m)~{\rm d}m,  \label{eq4}
\end{equation}
where $m_2$ is the mass of the SSP's MS turn-off point and $m_1$ is the
mass on the MS 2~mag below the turn-off point.

{\bf 3)} The SSP's $N_{\rm BSS}$ then follows from Eq.~(\ref{eq1}).

{\bf 4)} The distribution of the BSS population in the CMD is generated
using Monte Carlo simulations in two dimensions in the CMD, i.e.,
$M_V$ and $(B-V)$, based on the gaussian profiles described by
Eq.~(\ref{eq2}) and Table~\ref{table2}.

\begin{figure}
\includegraphics[width=6cm]{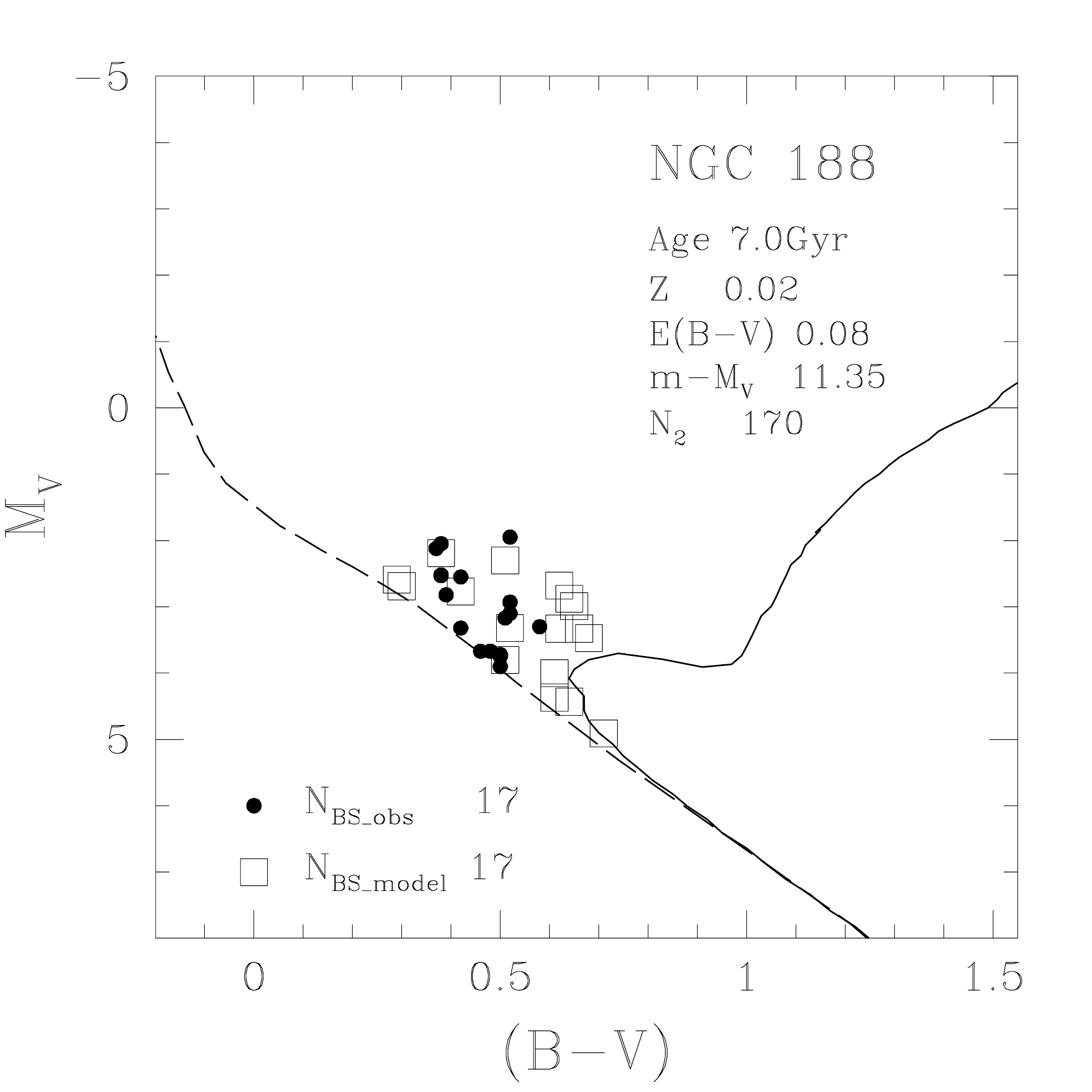}\includegraphics[width=6cm]{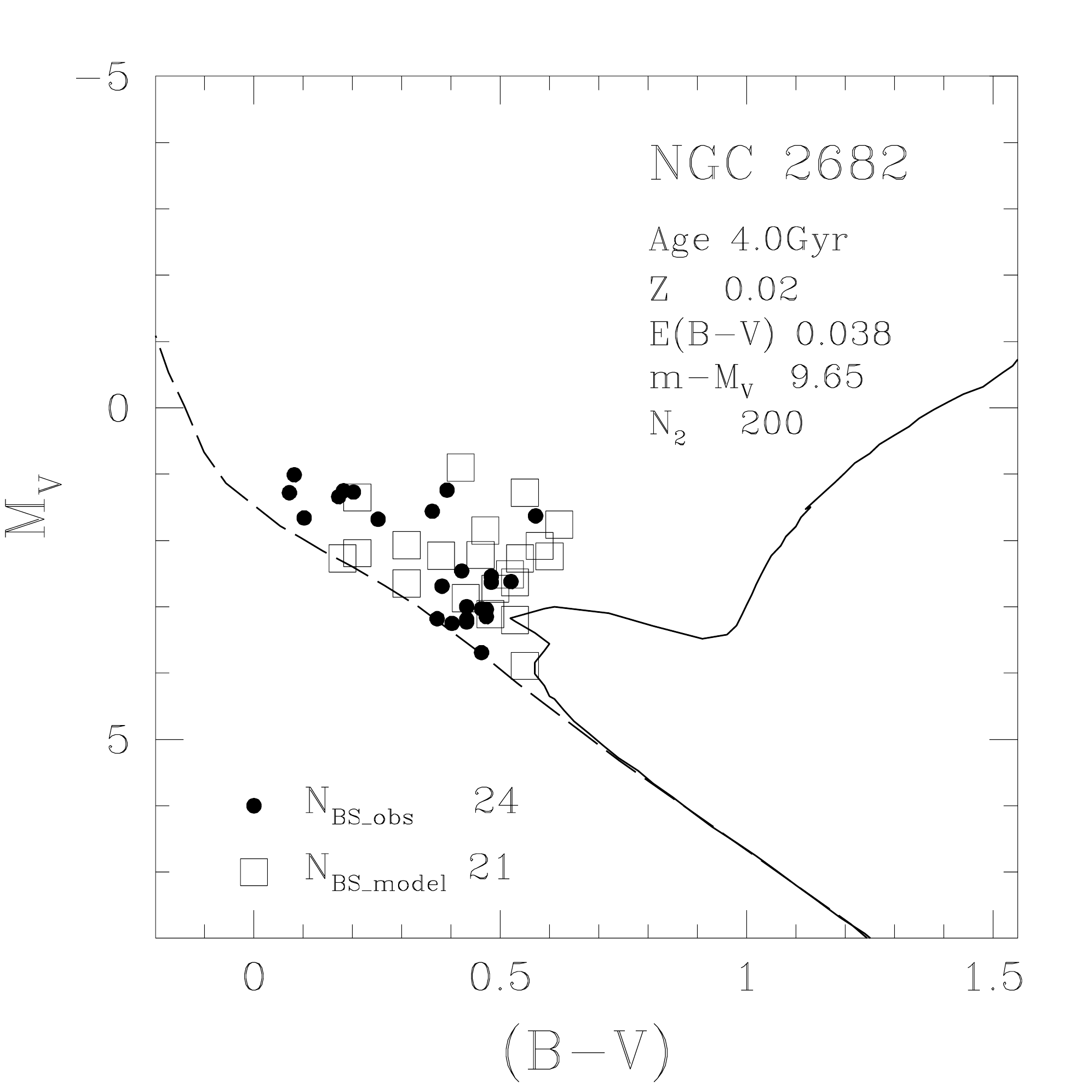}\\
\includegraphics[width=6cm]{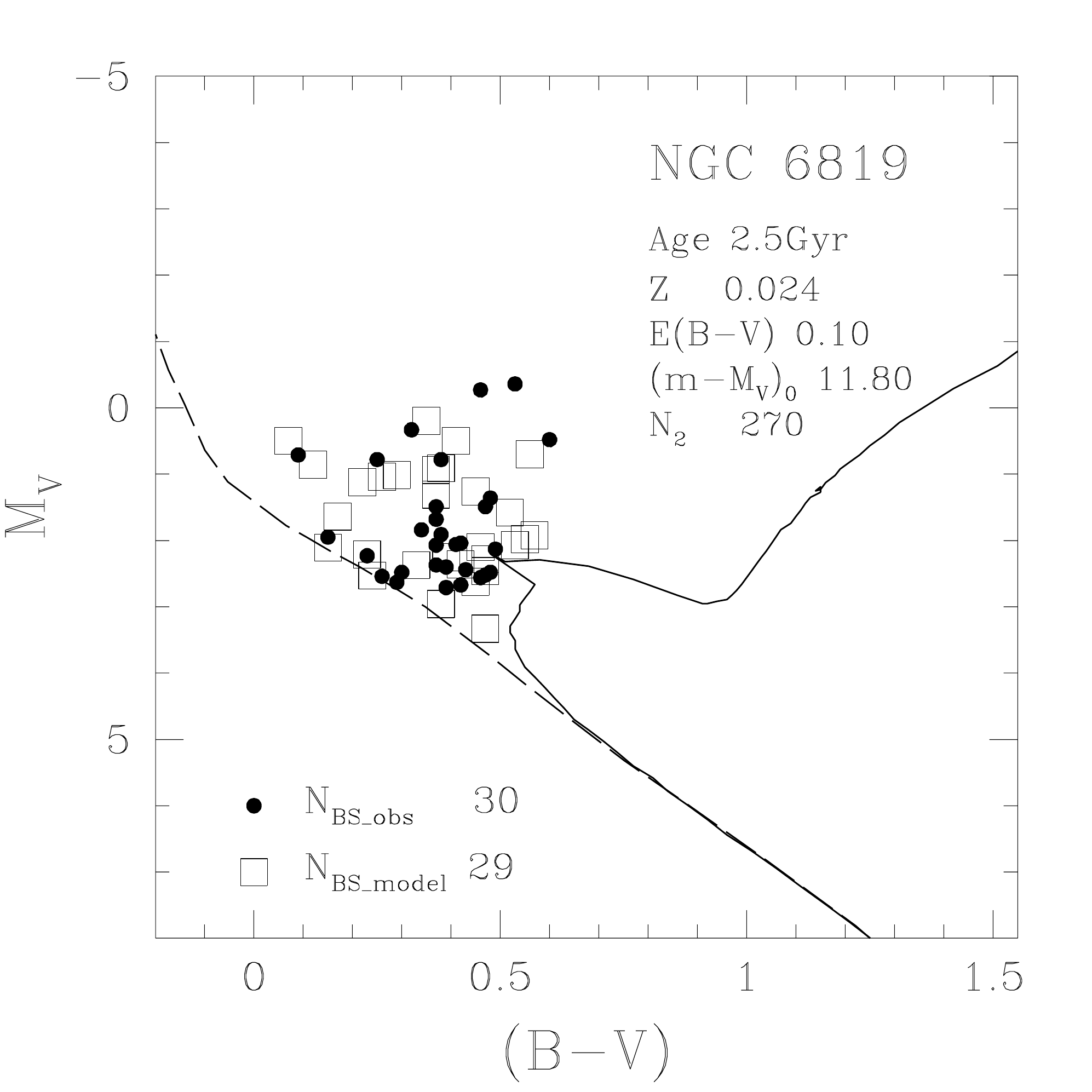}\includegraphics[width=6cm]{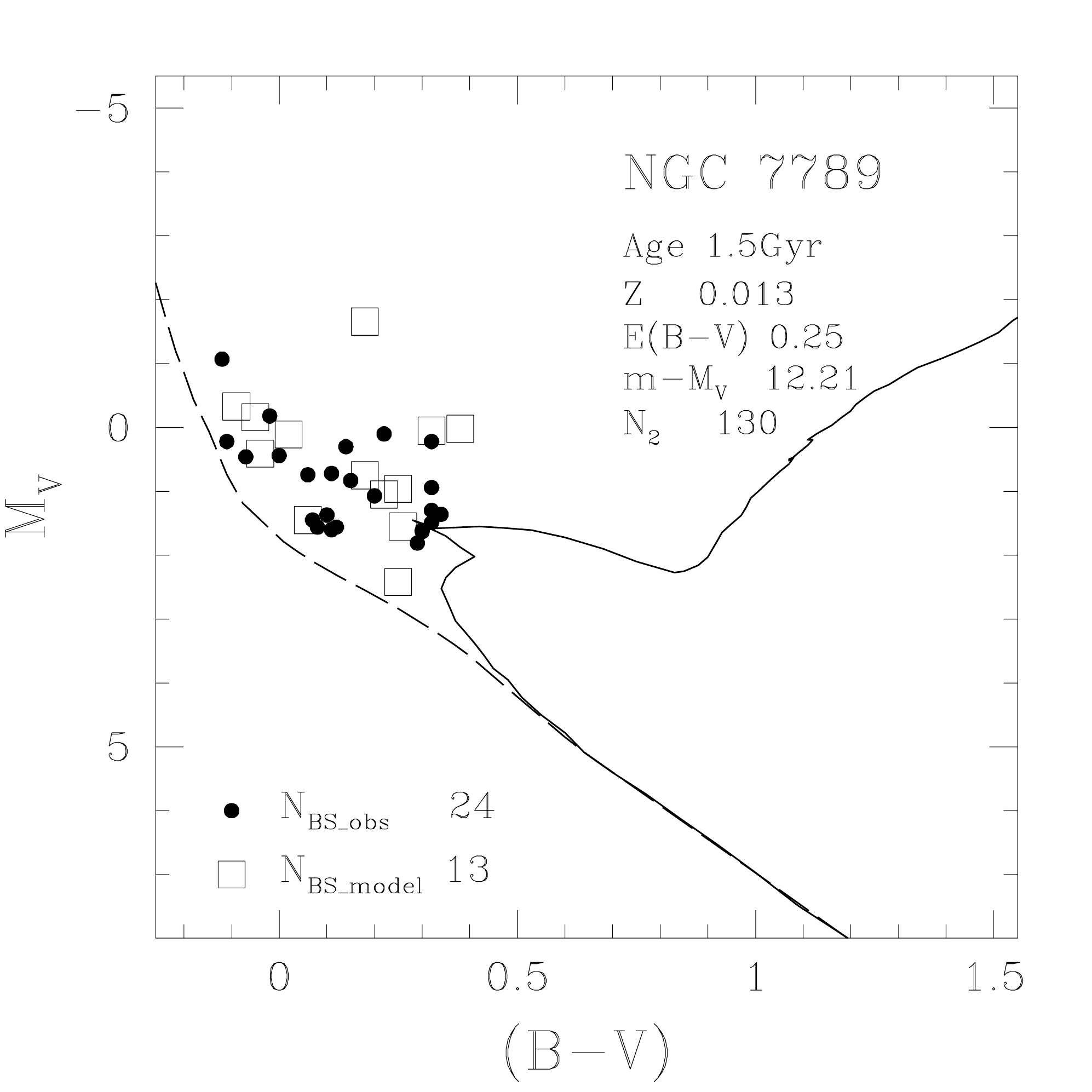}
\caption{Comparison between the observed and modelled BSS populations
in the CMDs of four representative Galactic OCs. In each panel, the
fundamental cluster parameters are given in the top right-hand
corner, the solid curve is the isochrone representing the cluster's
age and metallicity, the dashed line is the zero-age MS\index{ZAMS}, the solid
circles are the observed BSSs from AL95, and the open squares are the
model BSSs based on Monte Carlo simulations\index{Monte-Carlo simulation}. $N_{\rm BSS\_model}$ is
calculated using the observed value of $N_2$ in Eq.~(\ref{eq1}).}
\label{fig6}
\end{figure}

Figure~\ref{fig6} shows the comparison between the observed and modelled
BSS populations in the CMDs\index{colour-magnitude diagram} of four representative Galactic OCs\index{open cluster}. In
each panel, the fundamental cluster parameters are listed in the top
right-hand corner, the solid curve is the Padova~1994 isochrone for the
cluster's age and metallicity, the dashed line is the ZAMS, the solid
circles are the observed BSSs from AL95 and the open squares are the
BSSs generated by the Monte Carlo simulation. We calculate $N_{\rm
  BSS\_model}$ from the observed $N_2$ and Eq.~(\ref{eq1}). Apparently,
the BSS distribution fluctuates significantly for different simulations
because of the small $N_{\rm BSS\_model}$ numbers. What we intend to
show with this figure is that the modelled BSS population is quite
reasonable and comparable with that observed.

{\bf 5)} Calculate the spectrum of the SSP's BSS population.

For an SSP initially consisting of $10^5$ stars, $N_{\rm BSS}$ always
spans the range from dozens to at most hundreds of stars for ages
between 0.1 and 20~Gyr, which still gives rise to stochastic
fluctuations when we generate the BSS population in the CMD. These
fluctuations can influence the stability of the spectrum of the BSS
population. One of the best ways to reduce this stochastic effect is
by using the average of a large number of models. To find the optimum
number of realisations, \emph{(i)} we repeat the generation of BSS populations
in the CMD for a given SSP 500 times (500 models) and we subsequently
calculate the corresponding composite spectrum of the SSP. This
implies that the 500 models yield 500 different spectra for the same
SSP. \emph{(ii)} We calculate the broad-band colours $(U-B)$, $(B-V)$ and
$(V-R)$ for the SSP based on the average of the spectra 
of a successively increasing number till 500. This
calculation results in 500 values for each colour. \emph{(iii)} We calculate
the average of each colour using the colour values for between 200 and
500 models, thus quantifying the differences ($\delta$) of the 500
colour values and the average colour. Finally, \emph{(iv)} we identify the
number of models adopted to repeat generating the BSS population in the
CMD as the number for which $\delta < 0.0005$~mag. Using the
solar-metallicity SSPs as templates, we conclude that combining 100
models is a safe choice for SSPs characterised by different ages and
IMFs. 

\begin{figure}
\sidecaption
\includegraphics[width=7.8cm]{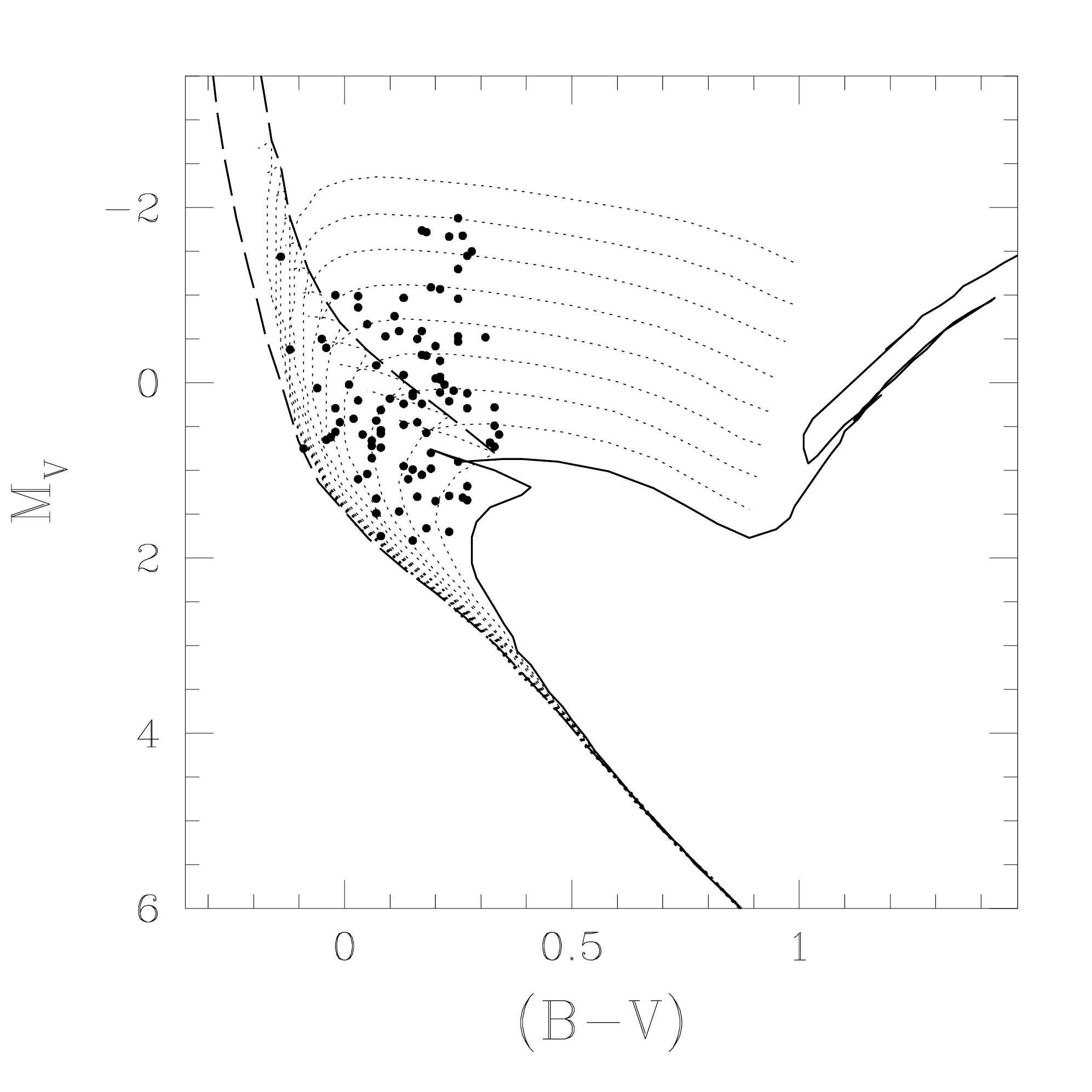}
\caption{Derivation of the basic parameters of BSSs in the CMD of a given cluster. The
  solid dots are the model BSSs, the solid curve is the isochrone for
  the SSP's age and metallicity corresponding to the cluster, and the dotted lines are isochrones
  representing ages younger than the SSP and truncated at the bottom
  of the RGB phase, that determine the fitted parameters and spectra of the BSSs. The dashed lines are the ZAMS and boundary between
  the MS and  the end of the MS stages, respectively.}
\label{FittingBS}
\end{figure}

To construct the spectrum of the model BSS population, we use
Padova~1994 isochrones of the same metallicity but younger ages than
the SSP to fit the position of each BSS in the CMD. We then derive the
effective temperature ($T_{\rm eff}$) and surface gravity (log~$g$) by
interpolation between two isochrones straddling the BSS. A
demonstration of this procedure is included in Fig.~\ref{FittingBS}. 
The solid dots are the model BSSs, the solid line is the isochrone for the
SSP's age and metallicity and the dotted lines are isochrones with
ages younger than the SSP and truncated at the bottom of the RGB
phase. The dashed lines are the ZAMS and the boundary of the MS and
post-MS stages, respectively. In our model construction, BSSs located
between the dashed lines are modelled strictly assuming that they can
be represented by the MS phases of the isochrones and the remainder of
the BSSs located outside the boundary are fitted with post-MS (and
pre-RGB) phases.

Depending on the values of $T_{\rm eff}$ and log~$g$, a spectrum is
extracted from the Lejeune et al. \cite{Lejeune+97} spectral library and assigned
to the BSS. The flux of the BSS spectrum is then calibrated using the
BSS's absolute magnitude and, finally, the spectrum of the BSS
population is obtained by adding up all flux-calibrated BSS spectra,
i.e., $F_{\rm BSS} = \sum_{i=1}^{N_{\rm BSS}}f_{\rm BSS}^i$.

The approximation to use the spectra of single MS stars to represent
the spectra of BSSs is made based on both theoretical (e.g., \cite{BH87,BH92}) and observational (e.g., \cite{Liu+08,SS00}) considerations. The major formation scenarios of BSSs,
such as mergers of primordial binaries and dynamical encounters
between stars, can replenish fresh hydrogen fuel in the core and
rejuvenate BSSs to the MS stage. Liu et al. \cite{Liu+08} studied the spectral
properties of a complete sample of 24 BSSs in M67\index{M67}
based on spectroscopic observations with a resolution of
3.2~{\AA}~pixel$^{-1}$ and covering wavelengths of 3600--6900
{\AA}. They concluded that BSS spectra can be well represented by the
theoretical spectra of single stars, at least at medium resolution.

{\bf 6)} We use BC03 models to represent the spectrum of the population of
normal member stars in an SSP (i.e., all member stars except the BSSs).

For a conventional SSP of age $t$ and metallicity $Z$, the integrated
spectrum is given by
\begin{equation}
 F_{\rm SSP} (\lambda, t, Z) = B \times \int_{m_{\rm l}}^{m_{\rm u}}
 \phi(m) f(\lambda, m, t, Z)~{\rm d}m, \label{eq5}
\end{equation}
where $\phi(m)$ is the IMF, $f(\lambda, m, t, Z)$ is the spectrum of a
single star of mass $m$, age $t$ and metallicity $Z$, $m_{\rm u}$ and
$m_{\rm l}$ are the upper and lower integration limits in mass,
respectively, and `$B$' is the normalisation constant required to
restore the real intensity of the flux of the conventional SSP.

\begin{figure}
\includegraphics[width=12cm]{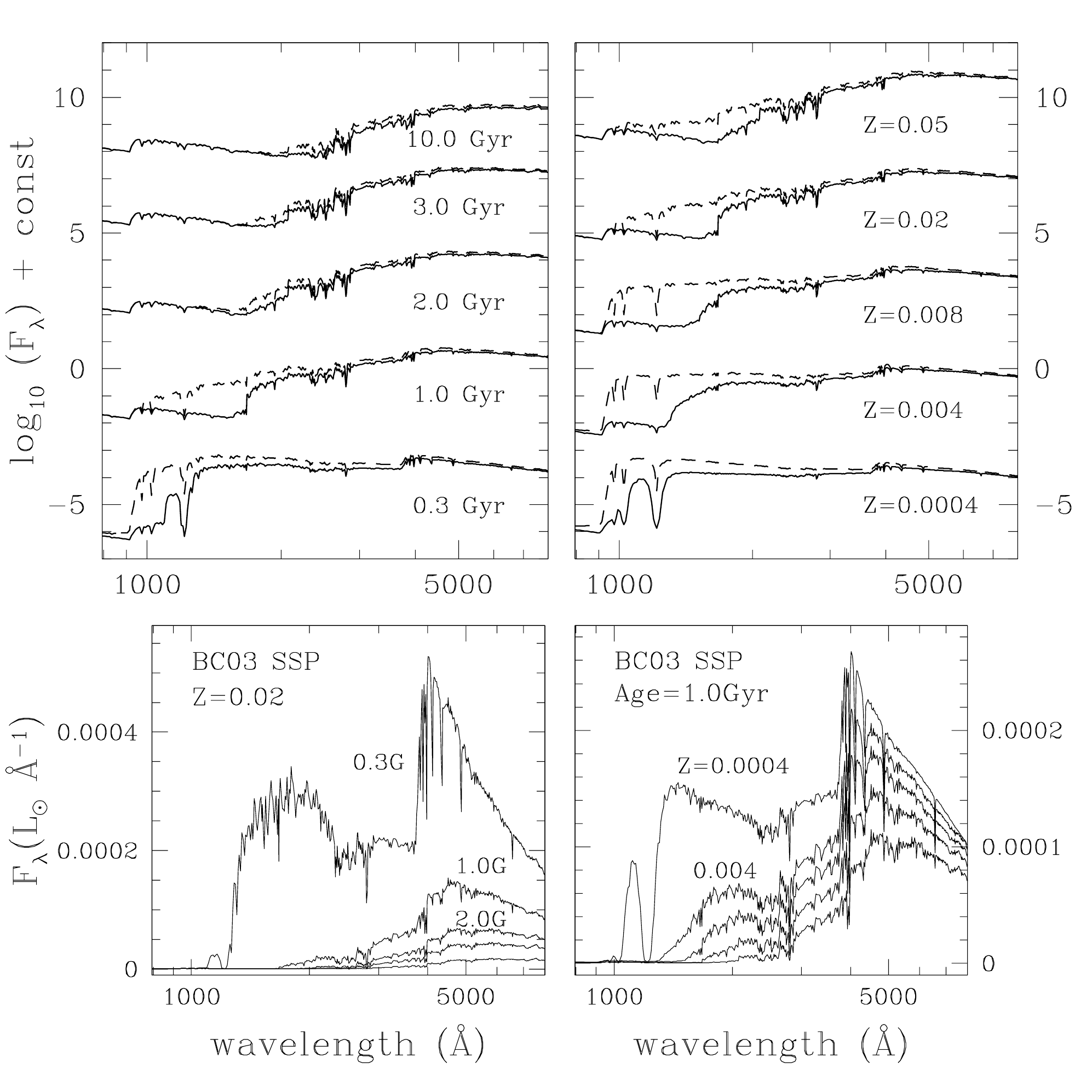}
\caption{Differences of the integrated spectral energy distributions
  (ISEDs) between BC03 and our models for different ages (top
  left-hand panel) and metallicities (top right-hand panel). The solid
  lines in the top panels are the BC03 ISEDs, while the dashed lines
  are the ISEDs resulting from our models. Because the BSS contribution
  is calculated as an increment to the BC03 models, the corresponding
  BC03 ISEDs are shown in the bottom panels for reference.}
\label{BC03Comp}
\end{figure}

Since BC03 normalised the total mass of their model SSPs to
1~M$_{\odot}$ at $t=0$, `$B$' is the total mass of the model SSP
containing $10^5$ stars at $t=0$:
\begin{equation}
 M_{\rm tot} = B = A \times \int_{m_1}^{m_2} \phi(m) m~{\rm d}m, \label{eq6}
\end{equation}
where `$A$' is the normalisation constant from Eq.~(\ref{eq3}), $m_1
=0.1$ and $m_2=100 {\rm M}_{\odot}$.

Similarly as for the calibration of the BSS spectra, the flux of the
conventional SSP is also calibrated based on its absolute
magnitude. $M_{\rm tot}$ is used to calculate $M_V$ of the
conventional SSP ($M_{V\_{\rm SSP}}$) based on Eq.~(\ref{eq7}), in
which $M_{V\_ {\rm SSP(BC03)}}$ is the absolute magnitude of the
corresponding BC03 SSP\footnote{$M_{\rm tot}$ actually quantifies the
relative increase in $M_V$ with respect to the $M_V$ of an initially 1~M$_{\odot}$ SSP.}:

\begin{equation}
M_{V\_{\rm SSP}} = M_{V\_{\rm SSP(BC03)}} + 2.5 \times \log_{10}
(1/M_{\rm tot}) \label{eq7}
\end{equation}

{\bf 7)} After flux calibration, direct combination of the spectra of the
BSS population and the conventional SSP yields the spectrum of the
BSS-corrected SSP. As our final step, the composite spectra are
normalised to the BC03 models by adopting the flux-calibration
constant derived from Eq.~(\ref{eq7}) for each SSP.

As an SSP ages, the SSP's BSS population evolves to redder colours
in the CMD following the movement of the SSP's MS turn-off, which
implies that the ``blue'' in ``blue stragglers'' only means ``bluer than
the MS turn-off'', not necessarily blue in colour. Since young SSPs
(age$<$0.1 Gyr) are not included in our models, and because of the
presence of ``yellow stragglers'' in the AL95 catalogue, we decided to
explore the BSS contributions to the integrated-light properties
involving $U$-, $B$- and $V$-band energies, because they may all be
significant. Specifically, we present and discuss in detail the
differences in the ISEDs,
broad-band colours and mass-to-light ratios ($M/L_V$) between our
models and those published by BC03 in this section.

Using the $Z=0.02$ models with a Salpeter IMF as example, the
differences in the ISEDs between two models of different age (top
left-hand panel) and metallicity (top right-hand panel) are given in
Fig.~\ref{BC03Comp}. The solid lines in the top panels represent the BC03
ISEDs, while the dashed lines represent the ISEDs from our
models. Since the BSS contribution is calculated as an increment to the
BC03 models, the intensities of the corresponding BC03 ISEDs of
different age (bottom left-hand panel) and metallicity (bottom
right-hand panel) are also given in the figure.

The differences presented in the top left-hand panel show a tendency
for a stronger BSS contribution for younger SSPs. BSSs have a
significant effect on ISEDs for ages between 0.1 and 1.0 Gyr. A sharp
enhancement appears in the UV\index{ultraviolet} range at 1.0 Gyr for our models, while
the UV intensity remains very low in the BC03 ISEDs. During SSP
evolution, all massive single stars have left the main sequence\index{main sequence star} and evolved into
the red supergiant or the red giant phases in conventional SSPs of
0.1--1.0 Gyr, the UV light declines and the near-infrared intensity
increases, and thus BSSs are the most luminous and bluest objects in
the populations. For SSPs older than 2.0 Gyr, the intensity of the BSS
contribution decreases smoothly and slowly as the population ages.
The bottom left-hand panel presents the BC03 ISEDs for different ages,
which show that ISEDs are stronger (brighter) for younger SSPs. This
tendency is consistent with that found for the BSS contribution to
ISEDs of different ages. 

\section{Discussions and Prospectives}\label{discuss}

Fitting the spectra of unresolved galaxies using SSP models to extract information of 
their star formation history and current stellar contents has been a routine practice in astrophysics, 
but stellar populations are never simple as the current theory of stellar evolution (basically of single stars) 
can predict. BSSs is one of the less understood type of stars, and they are probably 
one of the most important contributors to the uncertainties in EPS. 
The other chapters in this book addressed the complicated nature of individual BSSs in both observational 
and theoretical aspects. 

Based on observations of individual star clusters with determinations of BSSs in both numbers and 
photometric properties, we established a practical method to build ISED of these clusters and to use them 
as a constraint for the conventional SSP models and EPS analysis of galaxies.We have shown here the contents 
of BSSs actually altered the host stellar populations in terms of integrated light. 
Although there is still a very large uncertainties in the empirical treatment, our results show that EPS 
really should be built on more realistic SSP models that have all possible stars included properly. 
By analysing distribution of BSSs in a general composite CMD of all star clusters, we also developed a way 
to estimate BSS contributions in any stellar populations, and made a spectral library of BSS-SSP models 
that bares observationally constrained BSS content. The basic description of our BSS-SSP models includes:
\begin{enumerate}

\item The models cover the wavelength range from 91~{\AA} to 160
  $\mu$m, ages from 0.1 to 20 Gyr and metallicities $Z=0.0004, 0.004,
  0.008, 0.02$ (solar metallicity\index{metallicity}) and 0.05. The metallicity
  $Z=0.0001$ is not included, because extended horizontal branch
  stars\index{horizontal branch star}, instead of BSSs, dominate the energies in the UV and blue
  bands in such extremely metal-poor SSPs.

\item The models are constructed as increments to the BC03 standard
  SSP models using the Padova~1994 isochrones and the Lejeune et
  al. \cite{Lejeune+97} stellar spectra. They can thus be used directly in EPS
  studies as replacement of BC03 for the same parameter
  coverage. Application of the models should be limited to the
  ``low-resolution'' regime. As each BSS spectrum is approximated by the
  theoretical spectrum of a single MS star, the models cannot fully
  account for changes in the spectral lines that are related to the
  formation scenarios of BSSs.

\item The essential effect of BSSs is to make an SSP's ISED hotter in
  the UV, blue and optical bands, and consequently turn the broadband
  colours much bluer. Taking the SSP models with $Z=0.02$ as an
  example, the differences in the broad-band colours between BC03 and
  our models are 0.10$\pm$0.05~mag in $(U-B)$, 0.08$\pm$0.02~mag in
  $(B-V)$, 0.05$\pm$0.01~mag in $(V-R)$, 0.12$\pm$0.05~mag in $(u-g)$,
  0.09$\pm$0.02~mag in $(g-r)$ and 0.17$\pm$0.03~mag in $(g-z)$.
\end{enumerate}

Given the universal presence of BSSs in various stellar systems, the BSS-SSP models will enable the community 
to uncover interesting results in studies of stellar populations, although a number of limitations of the 
current set of models\footnote{available at: {\tt http://sss.bao.ac.cn/bss}}.
The main source of the uncertainties in our work is true BSSs in star clusters. 
The current catalog of BSSs in star clusters (AL95, AL07) actually came 
from observations taken with different instruments at different accuracy. 
A survey of star clusters producing a uniform and accurate catalog is needed. 
An update to the catalog can be expected in a near future \cite{Deng+13}. 
Also limited by small numbers of both BSSs and member stars, the statistics of BSS contribution to a cluster 
(population) is poor, this can be much improved by spectroscopy surveys of a substantially 
large number of individual stars (including BSSs) in our Galaxy , such as SDSS, LAMOST, Gaia, 
which can give direct census of BSSs in the Milky Way Galaxy.

\begin{acknowledgement}
This work is supported by National Science Foundation of China through grants No. Y111221001 and 10973015. We are also grateful for being invited to this ESO workshop, thanks to the organisers, in particular Giovanni Carraro and Henri Boffin.
\end{acknowledgement}

\backmatter
\printindex


\end{document}